\documentclass[11pt,a4paper]{article}
\usepackage[utf8]{inputenc}
\usepackage[T1]{fontenc}
\usepackage[english]{babel}
\usepackage[scale=0.8]{geometry}
\usepackage{mathtools}
\usepackage{amsmath,amssymb,amsthm}
\usepackage{graphicx}
\usepackage{cite}
\usepackage[dvipsnames,svgnames]{xcolor}
\usepackage{hyperref}
\usepackage[font=small,labelfont=bf]{caption}
\usepackage{appendix}
\usepackage{cancel}
\usepackage{ulem}
\usepackage{dsfont}
\usepackage{float}
\usepackage[final]{pdfpages}
\usepackage{setspace}
\usepackage{tcolorbox}
\newcommand{\ii}{\mathrm{i}}
\newcommand{\ee}{\mathrm{e}}
\newcommand{\dd}{\mathrm{d}}

\definecolor{pyblue}{RGB}{0,0,255}
\definecolor{pyfirebrick}{RGB}{178,34,34}

\begin{document}

\title{From agent-based dynamics to a kinetic theory of jellyfish swarms}

\author{Nicolas Perez$^{1,2}$, Erik Gengel$^1$, Zafrir Kuplik$^2$ and Eyal Heifetz$^1$}
\date{}

\maketitle

\begin{center}
    \textit{$^1$Department of Earth and Planetary Sciences, Tel Aviv University, 69978 Tel Aviv, Israel}\\[6pt]
    \textit{$^2$The Steinhardt Museum of Natural History, Tel Aviv University, 12 Klausner Street, 6901127 Tel Aviv, Israel}
\end{center}

\begin{center}
    email: \texttt{nicolasperez@tauex.tau.ac.il}
\end{center}

\begin{abstract}
    Massive jellyfish swarms observed at sea can extend over tens of kilometres and contain millions of individuals, yet the mechanisms governing their formation and large-scale dynamics remain poorly understood. Agent-based models provide a framework for describing this dynamics based on jellyfish responses to ocean currents and environmental cues, but become computationally prohibitive when extended to large populations and spatial scales relevant to ocean circulation. Here we derive a continuous kinetic theory from an active-particle model of jellyfish motion. The resulting Fokker-Planck framework incorporates transport by prescribed currents, stochastic reorientation, direct interactions and stimulated steering, allowing chemical signalling to be represented through a coupled field. We further derive a hydrodynamic closure for large swarms by exploiting the separation between fast orientational and slow spatial dynamics, yielding a reduced density equation suitable for implementation in ocean-current models. This framework provides a route from individual behavioural mechanisms to continuum descriptions of jellyfish populations and establishes a basis for constraining model parameters using observations and in-situ measurements. This approach offers a theoretical foundation for future numerical prediction of large jellyfish swarm formation and evolution in realistic ocean flows.
\end{abstract}

\section{Introduction}

\begin{figure}[h]
\begin{center}
    \includegraphics[scale=0.44]{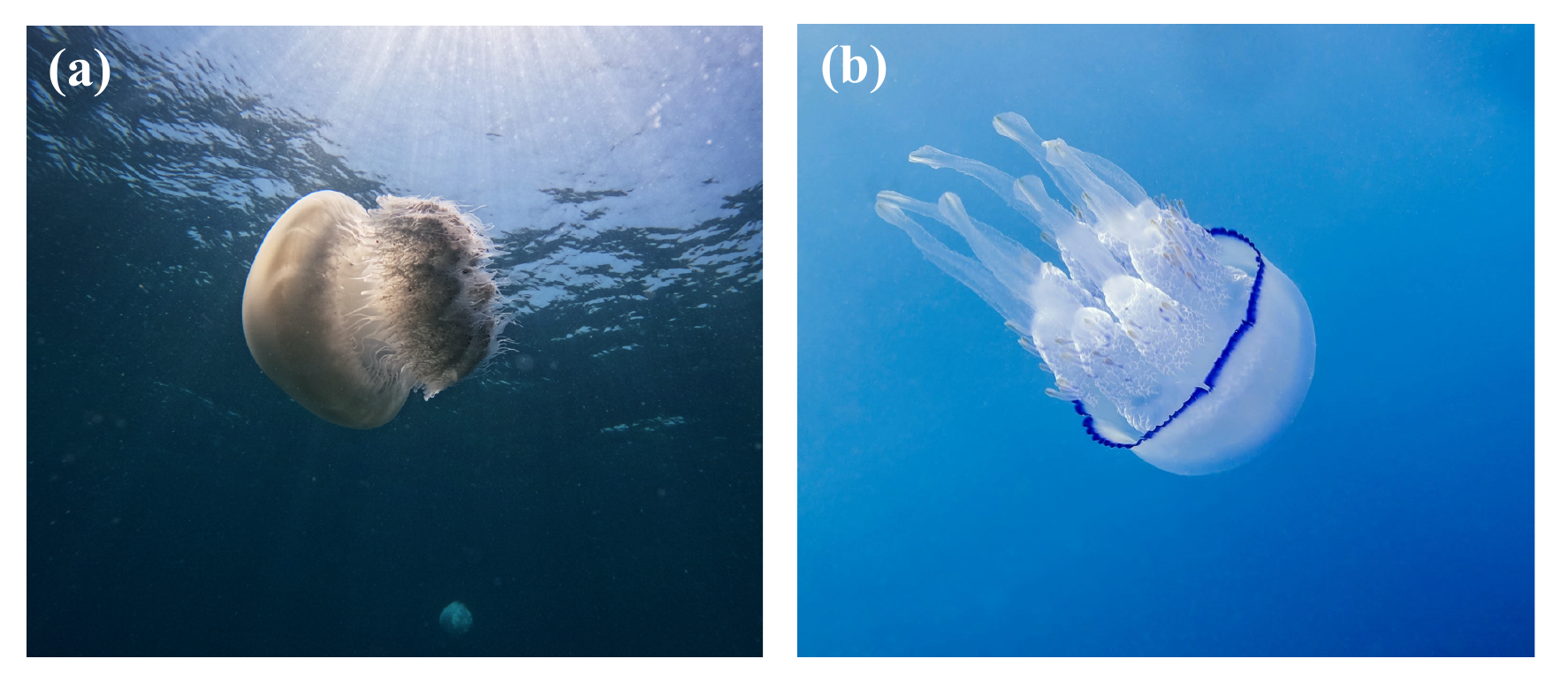}
\end{center}
\caption{\label{fig:jellyfish} Two swarming jellyfish species of the Mediterranean Sea, \textbf{(a)} \textit{Rhopilema nomadica} and \textbf{(b)} \textit{Rhizostoma pulmo}. Both species show propensity to swim against currents \cite{malul2019levantine}. Photo credits: Zafrir Kuplik \textbf{(a)} and Shevy Rothman \textbf{(b)}.}
\end{figure}

Large-scale aggregations of motile organisms are a widespread phenomenon in natural systems, emerging from the interplay between individual dynamics, environmental cues, and collective interactions \cite{ouellette2022physics}. In the marine environment, gelatinous zooplankton (jellyfish) are particularly prone to forming extended and concentrated patches that can persist over spatial scales ranging from kilometres up to tens of kilometres, even under weak interactions and challenging marine conditions \cite{douek2024long}. These swarms constitute striking examples of out-of-equilibrium structures in a fluid medium, characterised by strong spatial heterogeneity and pronounced temporal variability. Nevertheless, little is known about the origin of this phenomenon and its evolution, making it difficult to predict its occurrence, whose regularity and duration exhibit seasonal intermittency \cite{edelist2020phenological}. So far, most models employed to study swarm patterns have considered passive advection-diffusion dynamics, focusing on the role of marine circulation combined with strobilation hot spots \cite{edelist2022tracking,johnson2001developing,vodopivec2025spatially,dehos2026short}. However, in contrast with the common perception of jellyfish as passive drifters, recent studies have shown that several swarming species, for instance \textit{Rhopilema nomadica} and \textit{Rhizostoma pulmo} (Figure \ref{fig:jellyfish}), are actually efficient swimmers which seem to respond actively to environmental cues \cite{malul2019levantine}. Although little is known about the nature of such cues \cite{albert2011s} and how jellyfish respond to them, an outstanding feature is the ability and tendency of these jellyfish to swim against the local current \cite{malul2019levantine}, sometimes with speeds of the same order as typical sea current values (\textit{positive rheotaxis}). Although the biological mechanisms enabling jellyfish to infer current direction remain poorly understood, rheotactic behaviour is likely associated with a high sensitivity to fast, small-scale variations in the surrounding flow, which may confer functional advantages such as enhanced prey encounter and reduced onshore stranding \cite{albert2011s,fossette2015current,malul2019levantine}. Positive rheotaxis could thus be the main explanation behind certain peculiar patterns observed at sea, for instance collective offshore displacements against the direction of surface gravity waves \cite{malul2024directional}. This definitely suggests that passive models are neither sufficient to provide an accurate description of swarm motion, nor to explain the maintenance of such coherent structures at large scales.\\

To overcome the limitations of passive models, one must adopt a description of jellyfish's individual motion that includes swimming and active directional changes. This is the common approach of active matter models \cite{marchetti2013hydrodynamics}, which consist in describing the motion of living systems, such as bacteria \cite{schnitzer1993theory,saragosti2012modeling}, or self-propelled particles \cite{liebchen2018synthetic,tjhung2018cluster}, with laws that account for individual decision-making and response to environmental stimuli. Although simplified, such descriptions are often sufficiently accurate to capture the essence of certain collective phenomena such as aggregation, spontaneous long-range self-organisation \cite{toner1995long,bertin2006boltzmann} and motility-induced phase separation \cite{cates2015motility,soto2024kinetic}. Following this reasoning, Gengel \textit{et al} proposed an agent-based (Lagrangian) model describing jellyfish as two-dimensional \textit{active Brownian particles} (ABP), whose speed and orientation are continuously adjusted according to environmental factors that include sea currents and a hypothetical signalling mechanism \cite{gengel2023physics,gengel2024swarm}. This model naturally conveys the discussion of swarm persistence and evolution for jellyfish species with strong swimming abilities and sensibility to a variety of stimuli.\\

In this paper, we derive a continuum kinetic theory of jellyfish swarms, the coarse-grained formulation of the agent-based model of Gengel \textit{et al} \cite{gengel2023physics,gengel2024swarm} in terms of Eulerian fields. Our motivation is two-fold: 1/ As jellyfish swarms can be very large and dense, and sometimes consist of millions of individuals \cite{douek2024long}, a coarse-grained description appears to be a more effective framework than a Lagrangian one to study the phenomenon. 2/ Such a continuous kinetic theory allows us to regard swarm dynamics with a collective perspective, and address it using a hydrodynamic framework. The ultimate goal of this theory is to implement it in ocean-current models, both for research and forecasting purposes. In Section \ref{sec:agentbasedmodel}, we recall and briefly discuss the Lagrangian model of \cite{gengel2023physics,gengel2024swarm}, establish approximations that will be assumed in order to derive an adequate continuous kinetic theory, and introduce the Fokker-Planck framework. In Section \ref{sec:kineticmodel}, we express the Fokker-Planck equation that describes the coarse-grained version of the Lagrangian model, and provide a detailed analysis of each of its elements. In Section \ref{sec:hydrodynamics}, we analyse the different time scales and collective properties emerging from this kinetic formulation, and derive a hierarchy of hydrodynamic equations that covers the dynamics of jellyfish swarms under appropriate approximations, which can be implemented in ocean-current models for future studies. Finally, in Section \ref{sec:conclusion}, we summarise the results of this paper, and conclude by mentioning the main questions that we wish to address using the physical framework presented herein.

\section{From agent-based to continuum description of jellyfish motion} \label{sec:agentbasedmodel}

\subsection{Agent-based dynamics} \label{part:agentbasedmodel}

We recall here the mathematical description and general assumptions of the agent-based model proposed in \cite{gengel2023physics,gengel2024swarm}. As we aim to assess the importance of active swimming in shaping the large-scale  swarms' structure, we consider a two-dimensional model. Indeed, swarms can extend horizontally up to tens of kilometres while remaining comparatively shallow, up to a few tens of metres deep  for \textit{R. nomadica} \cite{gemmell2025movement,douek2024long}. Therefore, vertical motions are averaged out in the present context. Moreover, jellyfish are assumed to be identical and thus share the same properties, such as size, bell frequency, swimming velocity and response characteristics (while it is fair to assume that the active agents in a swarm consist of mature jellyfish of roughly same size, the average size itself exhibits seasonal variability \cite{edelist2020phenological}), introduced in the following. Two important additional assumptions are made about the mechanics of self-propulsion itself. First, we assume that the motion of each jellyfish is overdamped, i.e. that changes in (translational and rotational) velocity are immediate, which is a common assumption in active matter models. This means that jellyfish's inertia is neglected, as the force generated by self-propulsion compensates the external drag on their body. Second, jellyfish motion is affected by sea currents but the influence of bell motion on the flow is insignificant, so we consider that the currents are not affected by jellyfish. It is worth pointing out that this situation contrasts with microscopic active matter models in which moving bodies generate a low-Reynolds-number flow that transports them through active stress \cite{lushi2012collective}. In the present case, coupling with the flow (which has rather high Reynolds number) is unidirectional, since jellyfish are both transported and responsive to the currents, but do not affect them while swimming. Previous studies have suggested that swimming animals may contribute to fluid stirring and mixing \cite{Kwok2009,Katija2012}. However, to the best of our knowledge, such an effect has not yet been directly observed or quantified at the scale of jellyfish swarms.\\

\begin{figure}[h]
\begin{center}
    \includegraphics[scale=0.43]{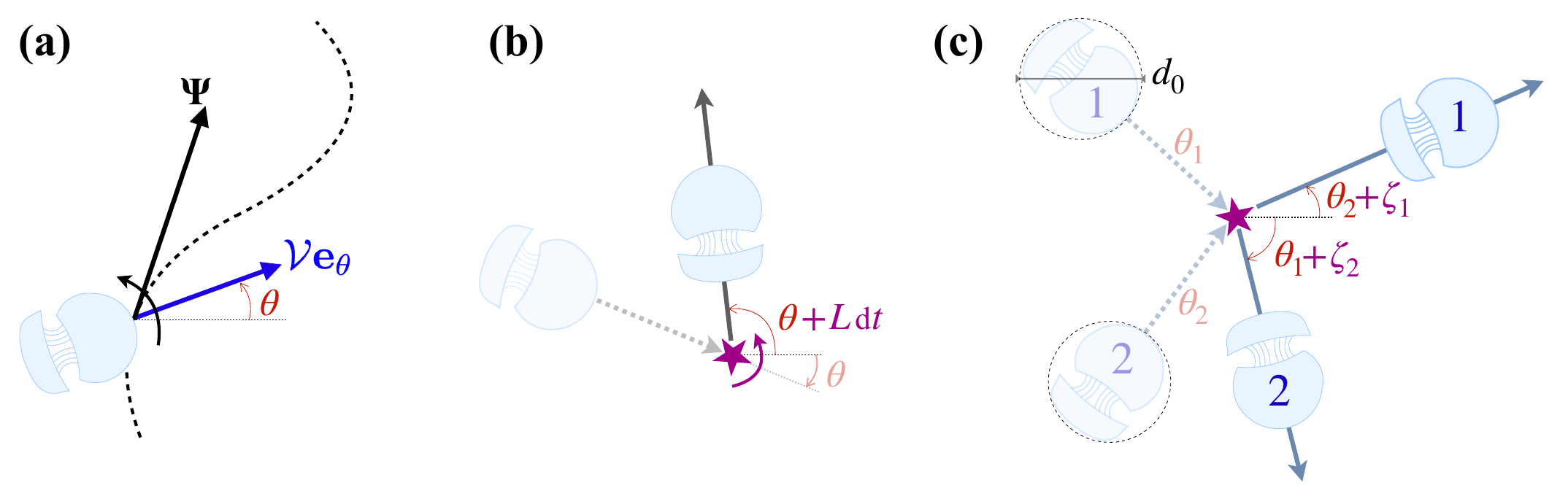}
\end{center}
\caption{\label{fig:process} Re-orientation of jellyfish occurs in different manners. \textbf{(a)} Continuous angular motion in response to different stimuli, represented by an external field $\boldsymbol{\Psi}$ which locally points towards most favorable conditions. \textbf{(b)} Random turning is included in the model as a representation of unpredictable behaviour and stochastic events such as turbulence. \textbf{(c)} Pairwise collisions occur when two individuals get closer than a distance $d_0$ representing the size of a jellyfish, at which point they both exchange orientation, up to an additional angle.}
\end{figure}

Each jellyfish $j$ at position $\mathbf{x}_j (t)$ on the $(x,y)$ plane, is both passively transported by the flow velocity $\mathbf{U}(\mathbf{x}_j,t)$, and actively swimming with speed $\mathcal{V}_j \mathbf{e}_{\theta_j}$, where $\mathbf{e}_{\theta_j} = \cos \left( \theta_j \right) \mathbf{e}_x + \sin \left( \theta_j \right) \mathbf{e}_y$ is the unit vector defining the orientation $\theta_j (t)$ of the jellyfish $j$ (Figure \ref{fig:process}). Gengel \textit{et al} \cite{gengel2024swarm} assume that the swimming amplitude $\mathcal{V}$ is a function of the local flow speed $|\mathbf{U}(\mathbf{x}_j,t)|$, modulated by a periodic function of time which embodies the pulsation of the bell:
\begin{equation} \label{eq:swimming_amplitude}
    \mathcal{V}_j (t) = \mathcal{V} \left( |\mathbf{U}(\mathbf{x}_j,t)| \right) \mathcal{B} (\varphi_j) \ ,
\end{equation}
where $\mathcal{B}$ is a $2 \pi$-periodic function and $\varphi_j (t) = \Omega t + \varphi_j^0$ is the bell phase of jellyfish $j$. The amplitude $\mathcal{V}$ is a generic increasing function of $|\mathbf{U}|$, which has finite limits as $|\mathbf{U}| \rightarrow 0$ and $|\mathbf{U}| \rightarrow +\infty$. The environmental factors affecting the swimming amplitude are not known, and the latter probably depends not only on the intensity of the current but also on other external stimuli and their gradient. Nevertheless, we adopt this hypothesis, since the central aspect of jellyfish collective dynamics does not lie in the exact form of $\mathcal{V}$ but rather in the dynamics of the orientation $\theta_j$, as we will demonstrate in the rest of this paper. Our model assumes that rotation is driven by different factors, that we can classify into three categories:

\begin{itemize}
    \item Deterministic, continuous rotational motion representing the adjustment of jellyfish's orientation in response to external stimuli (Figure \ref{fig:process}a).
    \item Individual stochastic turning (self-diffusion), which represents both jellyfish's inner degree of unpredictability and random external events owing, for instance, to small scale turbulence (Figure \ref{fig:process}b).
    \item Discrete turns owing to collisions between pairs of jellyfish (Figure \ref{fig:process}c).
\end{itemize}

Random turning is expressed with a torque $L_j(t)$ which is a stochastic process, while the response to environmental stimuli is represented as a deterministic mechanism. The neuronal net of jellyfish produces a dynamical reaction to the perception of stimuli, which is encoded in a function that couples the orientation $\theta$ to the amplitude and direction of the different stimuli. Since such a response function is generally $2 \pi$-periodic in $\theta$, Gengel \textit{et al} \cite{gengel2024swarm} retain the dominant order and thus assume a dipolar interaction term. Therefore, in the absence of collision with another jellyfish or an obstacle, $\theta_j$ evolves as
\begin{equation} \label{eq:thetadot}
    \dot{\theta}_j = L_j (t) - \sum_{s} \varepsilon_s \sin \left( \theta_j - \theta_s (\mathbf{x}_j) \right) \ ,
\end{equation}
where the sum is over all different external stimuli "$s$" considered, characterised by a local favorable direction $\theta_s$, to which the jellyfish adjusts its orientation at rate $\varepsilon_s$. Note that $\theta_s$ and $\varepsilon_s$ can be functions of time and position, via the external factors themselves. By summing over different stimuli, we assume that a jellyfish makes an average decision in the presence of a variety of possibly competing stimuli. The random process $L_j (t)$ is a colored noise (Ornstein-Uhlenbeck) with correlation rate $\lambda_\theta$ and angular diffusivity $D_\theta$:
\begin{equation} \label{eq:OUnoise}
    \dot{L}_j = -\lambda_\theta L_j + \sqrt{2 D_\theta} \ \eta_j (t) \ , \quad \text{with} \quad \langle \eta_j (t) \rangle = 0 \quad \text{and} \quad \langle \eta_i (t) \eta_j (t') \rangle = \delta_{ij} \delta (t-t') \ .
\end{equation}

Adopting the dipolar steering term of Expression \eqref{eq:thetadot} to characterise their response to external stimuli, we assume that jellyfish have enough sensory access to directly probe physical gradients of relevant stimuli such as food or signalling chemical (spatial sensing), owing to a relatively complex nerve net connected to their sensory organs (\textit{rhopalia}) \cite{graham2001physical,rakow2006orientation,albert2011s}. In contrast, microscopic organisms like \textit{E. coli} bacteria infer gradients indirectly by temporal sensing, i.e. by moving around the surrounding environment and turning randomly with lower frequency if the conditions are becoming more favorable -- a process called \textit{run-and-tumble} \cite{alt1980biased,schnitzer1993theory}.\\

Ideally, one needs to know exactly which stimuli individuals of a given jellyfish species are sensitive to and how they react to those stimuli, in terms of directional swimming, however, there is no general answer to this question and most hypotheses are still under investigation. Nevertheless, observations and measurements have shown that certain jellyfish species actively respond to numerous natural factors \cite{albert2011s}. Besides, some studies suggest that swarm formation may be strongly affected by oceanic fronts and thus marine-related physical gradients \cite{graham2001physical,rakow2006orientation,fossette2015current,manko2022oceanic,malul2024directional}. In particular, reference \cite{rakow2006orientation} suggests that swimming patterns of jellyfish swarms can be affected by sensitivity to shear, while studies like \cite{fossette2015current,malul2024directional} have demonstrated the significance of counter-current swimming (positive rheotaxis), a behaviour that has been observed in different species but whose biological origin is still unknown \cite{albert2011s,malul2019levantine}. To emphasise the interplay between jellyfish swarms and marine currents, Gengel \textit{et al} \cite{gengel2024swarm} focus on three orientational responses: positive rheotaxis, shear avoidance and chemical signalling. The dipolar steering term in Expression \eqref{eq:thetadot} is thus a sum of three terms, which can be alternatively written by introducing a \textit{steering vector field}
\begin{equation} \label{eq:vectorpsi}
    \boldsymbol{\Psi} = \sum_s \varepsilon_s \mathbf{e}_{\theta_s} \equiv \Psi \mathbf{e}_{\theta_\Psi} \ ,
\end{equation}
whose direction $\theta_\Psi$ indicates the local average preferred orientation of jellyfish, and whose norm $\Psi$ is the rate at which they align to this direction. From definition \eqref{eq:vectorpsi}, the deterministic orienting term of Equation \eqref{eq:thetadot} can be written
\begin{equation} \label{eq:Kuramoto_steering}
    - \sum_{s} \varepsilon_s \sin \left( \theta_j - \theta_s (\mathbf{x}_j) \right) = -\Psi \sin(\theta_j - \theta_\Psi ) \ .
\end{equation}

Among the different environmental cues considered, positive rheotaxis means aligning against the flow, i.e. with the local direction of $-\mathbf{U}$, shear avoidance is characterised by swimming towards zones of low magnitude of  vorticity, indicated by $-\boldsymbol{\nabla}\mathcal{C}$ with $\mathcal{C} = |\boldsymbol{\nabla} \times \mathbf{U}|$, and signalling attraction means that jellyfish orient themselves towards the concentration gradient of a hypothetical chemoattractant, $\boldsymbol{\nabla} \phi$. This chemoattractant is a passive tracer produced by jellyfish themselves and transported by sea currents, and thus obeys the following advection-diffusion equation:
\begin{equation} \label{eq:signalling}
    \partial_t \phi + \boldsymbol{\nabla} \cdot \left( \phi \mathbf{U} \right) - D_\phi \boldsymbol{\nabla}^2 \phi = S \ ,
\end{equation}
where the source term $S$ depends on the relative position of jellyfish $\mathbf{x}_j$ to the Eulerian variable $\mathbf{x}$, and the concentration $\phi$ itself (see appendix B in \cite{gengel2024swarm}). We propose in \ref{part:chemical_source} an expression for $S$, which is more straightforward in the Eulerian framework.\\

Let us note that the form of the dipolar steering term \eqref{eq:Kuramoto_steering} is equivalent to that of the Kuramoto model \cite{kuramoto2003chemical}, where $\Psi$ plays a role analog to the order parameter, since it measures directionality. However, in the Kuramoto model or the Vicsek model \cite{vicsek1995novel}, spontaneous collective self-organisation results from direct aligning interactions between individuals (e.g. visual, acoustic, etc. \cite{ouellette2022physics}). In contrast, we assume that the collective alignment of jellyfish is not based on direct agent-agent interactions but rather on external cues, which can include indirect chemotactic sensing (to this extent, swarm dynamics is more relatable to the Keller-Segel model \cite{keller1970initiation,keller1971model}, for instance). Interactions between jellyfish are limited to passive collisions that do not contribute to their collective organization. A mathematical treatment of collisions will be addressed in \ref{part:collision}.\\

The agent-based model presented above is meant to provide a plausible theoretical framework to study the phenomenon of jellyfish swarms. It relies on informed assumptions and not direct experimental measurements. Future experimental observations will be needed to validate this model and refine the variables introduced in Expressions \eqref{eq:swimming_amplitude} and \eqref{eq:thetadot}, i.e. $\mathcal{V}_j$, $L_j$, $\varepsilon_s$.

\subsection{Kinetic formulation}

Since jellyfish swarms can extend over tens of kilometres, their large-scale collective dynamics are likely slaved to mesoscopic marine current structures. Moreover, because such swarms may comprise millions of individuals, they may be reasonably described by continuous density fields obeying partial differential equations derived from the underlying agent-based model. This provides the starting point of our kinetic theory. One must therefore reformulate the Lagrangian model introduced in \ref{part:agentbasedmodel} in statistical terms, i.e. as a Fokker–Planck equation \cite{oksendal2013stochastic}, which is the ensemble-mean formulation of the model previously described.\\

For this purpose, we introduce the distribution function in configuration space, $f(\mathbf{x},\theta,t)$, which represents the number of jellyfish per unit of surface and orientation angle at position $\mathbf{x}$ and angle $\theta$, and time $t$. This function is similar to a density, keeping the angular information to include all degrees of motion of the jellyfish. The conservation equation of jellyfish number in this configuration space, in flux form, reads as
\begin{equation} \label{eq:flux-FK}
    \partial_t f + \boldsymbol{\nabla} \cdot \left( f \dot{\mathbf{x}} \right) + \partial_\theta \left( f \dot{\theta} \right) = I_\text{col} \left[ f,f \right] \ ,
\end{equation}
where $\boldsymbol{\nabla} \cdot$ is the divergence operator in position space (here, $\boldsymbol{\nabla} = \partial_x \mathbf{e}_x + \partial_y \mathbf{e}_y$). The LHS of Equation \eqref{eq:flux-FK} contains all terms characterizing the free motion of jellyfish, in both position (second term) and orientation (third term), as introduced in \ref{part:agentbasedmodel}. In contrast, the RHS represents non-conservative events, namely pairwise collisions (Figure \ref{fig:process}c), characterised by a nonlinear (quadratic) operator $I_\text{col} \left[ f,f \right]$. The next section is dedicated to the explicit derivation of all elements contained in Equation \eqref{eq:flux-FK}.

\section{Continuum kinetic model} \label{sec:kineticmodel}

\subsection{Translational and orientational dynamics}

In virtue of the Lagrangian model described in \ref{part:agentbasedmodel}, velocity in positional space is the sum of passive transport and self-propulsion, i.e.
\begin{equation} \label{eq:agent_velocity}
    \dot{\mathbf{x}} = \mathbf{U} + \mathcal{V} \mathbf{e}_\theta \ .
\end{equation}

As explained in \ref{part:agentbasedmodel}, the amplitude of jellyfish velocity is assumed to depend on the local value of $|\mathbf{U}|$ only, up to a periodic factor $\mathcal{B}$ representing the modulation of the swimming velocity over a bell cycle. Since the respective bell phase of each jellyfish is uncorrelated to the others and the period of bell pulsation is typically of order one second, which is very short compared to the relevant time scales that define swarming events, this bell pulsation term is averaged over a period, and the mean amplitude itself is noted $\mathcal{V}$ in the rest of the paper. Since $\mathcal{V}$ is a function of flow amplitude, jellyfish velocity \eqref{eq:agent_velocity} (passive transport and self-propulsion) is thus fully defined by position $\mathbf{x}$ (through $\mathbf{U}$ and $\mathcal{V}$) and orientation $\theta$ (through $\mathbf{e}_\theta$).\\

As for the third term of the LHS of Equation \eqref{eq:flux-FK}, the evolution of $\theta$ is driven by the deterministic dipolar steering and the stochastic torque $L(t)$ of Equation \eqref{eq:thetadot}:
\begin{equation} \label{eq:agent_rotation}
    \dot{\theta} = - \Psi \sin \left( \theta - \theta_\Psi \right) + L \ .
\end{equation}

In this study, we consider the collective dynamics on time scales long compared with the characteristic period of the bell and jellyfish memory, represented for instance by the correlation time $\lambda_\theta^{-1}$ of the stochastic torque $L$. Although there are no direct measurements yet to reliably determine $\lambda_{\theta}$, this correlation time is indeed evaluated to be very short (Gengel \textit{et al} \cite{gengel2024swarm} assume $\lambda_{\theta}^{-1} \approx 0.2$ s). We thus coarse-grain the Ornstein–Uhlenbeck process by approximating the angular velocity $L$ by $\sqrt{2 D_\theta/\lambda_\theta^2} \ \eta(t)$, i.e. as an effective white-noise process with rotational diffusivity $D_r = D_\theta/\lambda_\theta^2$. The latter yields a diffusion term \cite{risken1989fokker}, therefore, putting together \eqref{eq:flux-FK}, \eqref{eq:agent_velocity} and \eqref{eq:agent_rotation}, $f$ obeys the Fokker-Planck equation
\begin{equation} \label{eq:Boltzmann}
    \partial_t f + \boldsymbol{\nabla} \cdot \left( f \left( \mathbf{U} + \mathcal{V} \mathbf{e}_\theta \right) \right) - \partial_\theta \left( f \Psi \sin \left( \theta - \theta_\Psi \right) \right) = D_r \partial_{\theta \theta} f + I_\text{col} \left[ f,f \right] \ .
\end{equation}

By definition, the variables $\mathbf{U}, \mathcal{V}, \Psi$ and $\theta_\Psi$ in Equation \eqref{eq:Boltzmann} are independent of orientation $\theta$. It is instructive to emphasise the contrast between this model and the Vicsek model \cite{vicsek1995novel}, as Equation \eqref{eq:Boltzmann} is similar to Equation (1) in \cite{bertin2006boltzmann}, yet with the following important nuances:

\begin{itemize}
    \item The second term in the LHS of \eqref{eq:Boltzmann} accounts for both passive transport ($\mathbf{U}$) and self-propulsion ($\mathcal{V} \mathbf{e}_\theta$), and the swimming amplitude $\mathcal{V}$ can vary according to environmental conditions, while the Vicsek model considers $\mathbf{U}=0$ and constant $\mathcal{V}$.
    \item Angular self-diffusion -- the first term in the RHS of \eqref{eq:Boltzmann} -- is here modelled as a continuous stochastic process, instead of a discrete one as in \cite{bertin2006boltzmann}, although diffusion is indeed the continuous limit of a discrete Gaussian jump process: with the notation of \cite{bertin2006boltzmann}, $D_r = \lambda \sigma^2 / 2$ as $\lambda$ becomes very large and $\sigma$ very small (see for instance \cite{gardiner2009stochastic}).
    \item More importantly, in the Vicsek model, collective spontaneous self-organisation (\textit{flocking}) emerges owing to the interaction term, $I_\text{col}$, which depicts individuals actively aligning with each other. In contrast, observations of jellyfish behaviour suggest no direct interactions, and directional motion is instead driven by local external stimuli, represented by the third term in the LHS of \eqref{eq:Boltzmann}. The effect of collisions is difficult to specify at the inter-individual level, as the response of jellyfish to steric encounters is not sufficiently characterized and no clear mechanism has been established. Nevertheless, it is reasonable to assume that collision events have a passive effect on jellyfish motion (they simply "bump into each other"). A mean-field expression for this mechanism is proposed in the next subsection.
\end{itemize}

\subsection{Density-dependent angular diffusion from pairwise interactions} \label{part:collision}

Gengel \textit{et al} \cite{gengel2024swarm} use soft-repulsion interaction potentials to represent direct collisions (volume exclusion), however, expressing the latter as a scattering integral $I_\text{col}[f,f]$ (keeping only binary collisions \cite{bertin2006boltzmann}) for the Fokker-Planck formulation \eqref{eq:Boltzmann} is not a simple task \cite{ernst2006boltzmann,gallagher2013newton} and, in general, the result is the integral of a complex function that depends on the angles and impact parameter. As previously mentioned, direct collisions in jellyfish swarms are likely passive and do not have a significant effect, except in very dense swarms where jellyfish are effectively unable to sustain directional swimming over large distances. Therefore, we must adopt a model for pairwise collisions that effectively reduces the mean jellyfish velocity as density increases. Instead of soft repulsion, we model the coarse-grained effect of binary collisions as an additional density-dependent angular diffusivity, $\delta(\rho)$, meaning that particles in dense regions lose directional persistence faster because they collide more frequently, which increases their angular decorrelation rate. Besides, with passive hard or soft repulsion, particle velocity is generally not conserved after a collision event, yet we assume here that the swimming speed $\mathcal{V}$ only depends on external parameters. Therefore it seems more consistent to adopt a model in which jellyfish change orientation but keep swimming at the same speed $\mathcal{V} (\mathbf{x},t)$ immediately after a collision at time $t$ and position $\mathbf{x}$. It is the same point of view as adopted in \cite{vicsek1995novel,bertin2006boltzmann} for flocks. Given the relatively dilute nature of the aggregations considered here, we therefore adopt a minimal mean-field closure in which collisions contribute an additional density-dependent angular diffusion, assumed to increase with local density:
\begin{equation} \label{eq:collision}
        I_\text{col} \left[ f,f \right] = \delta(\rho) \partial_{\theta \theta} f \ ,
\end{equation}
where $\delta(\rho) = \nu d_0 \mathcal{V} \rho$ ($\nu$ is a dimensionless steric factor that depends on jellyfish shape and the nature of collisions), $d_0$ is the diameter of jellyfish, and
\begin{equation} \label{eq:density}
    \rho (\mathbf{x},t) \equiv \int_{-\pi}^{\pi} f(\mathbf{x},\theta,t) \, \dd \theta
\end{equation}
is the density of jellyfish, summing over all possible orientations. Appendix A proposes a derivation of Expression \eqref{eq:collision} as a simplified approximation of a more general collision integral. This mean-field model is consistent with the basic properties of the Fokker-Planck equation itself (e.g. Expression \eqref{eq:collision} preserves the number of particles), and yields the main expected effect of collisions on collective motion, namely to increase disorientation as jellyfish aggregate, thus eventually saturating the aggregation mechanism. This manifests as a total angular diffusivity
\begin{equation} \label{eq:D}
    \mathcal{D} (\rho) = D_r + \delta(\rho) \ ,
\end{equation}
which increases with $\rho$. Besides, this additional diffusivity $\delta$ naturally increases with jellyfish size ($d_0$) and mobility ($\mathcal{V}$). In the present study, we consistently assume that $\delta$ remain small or comparable with $D_r$, i.e. that collision events do not dominate rotational self-diffusion. Under this formulation, the quantity $\mathcal{V}/\mathcal{D}$ is the local mean free path of jellyfish, i.e. the average distance travelled by a jellyfish between two events of stochastic nature (either self-diffusion or collision).\\

Expression \eqref{eq:collision} yields no active structuring effect, unlike binary collisions in the theory of motility-induced phase separation, where they induce a net displacement that can lead to an effective negative diffusivity \cite{cates2015motility,soto2024kinetic}. Usually, jellyfish giant swarms remain dilute enough so their aggregation is likely not motility-induced but rather emerges from collective response to external cues. Because experimental constraints are currently unavailable, we adopt this minimal mean-field model, keeping in mind that all the parameters and variables of this model are meant to be tested using experimental data. Equation \eqref{eq:Boltzmann} can thus be written in the final form
\begin{equation} \label{eq:Fokker-Planck}
    \partial_t f + \boldsymbol{\nabla} \cdot \left( f \left( \mathbf{U} + \mathcal{V} \mathbf{e}_\theta \right) \right) = \partial_\theta \left( f \Psi \sin \left( \theta - \theta_\Psi \right) \right) + \mathcal{D}(\rho) \partial_{\theta \theta} f \equiv \mathcal{L}_\theta \left[ f \right] \ ,
\end{equation}
where $\mathcal{L}_\theta$ is a linear angular operator whose coefficients depend on position $\mathbf{x}$ and time $t$ through $\Psi, \theta_\Psi$ and $\mathcal{D}(\rho)$.

\subsection{Chemical source and signalling} \label{part:chemical_source}

\begin{figure}[h]
\begin{center}
    \includegraphics[scale=0.35]{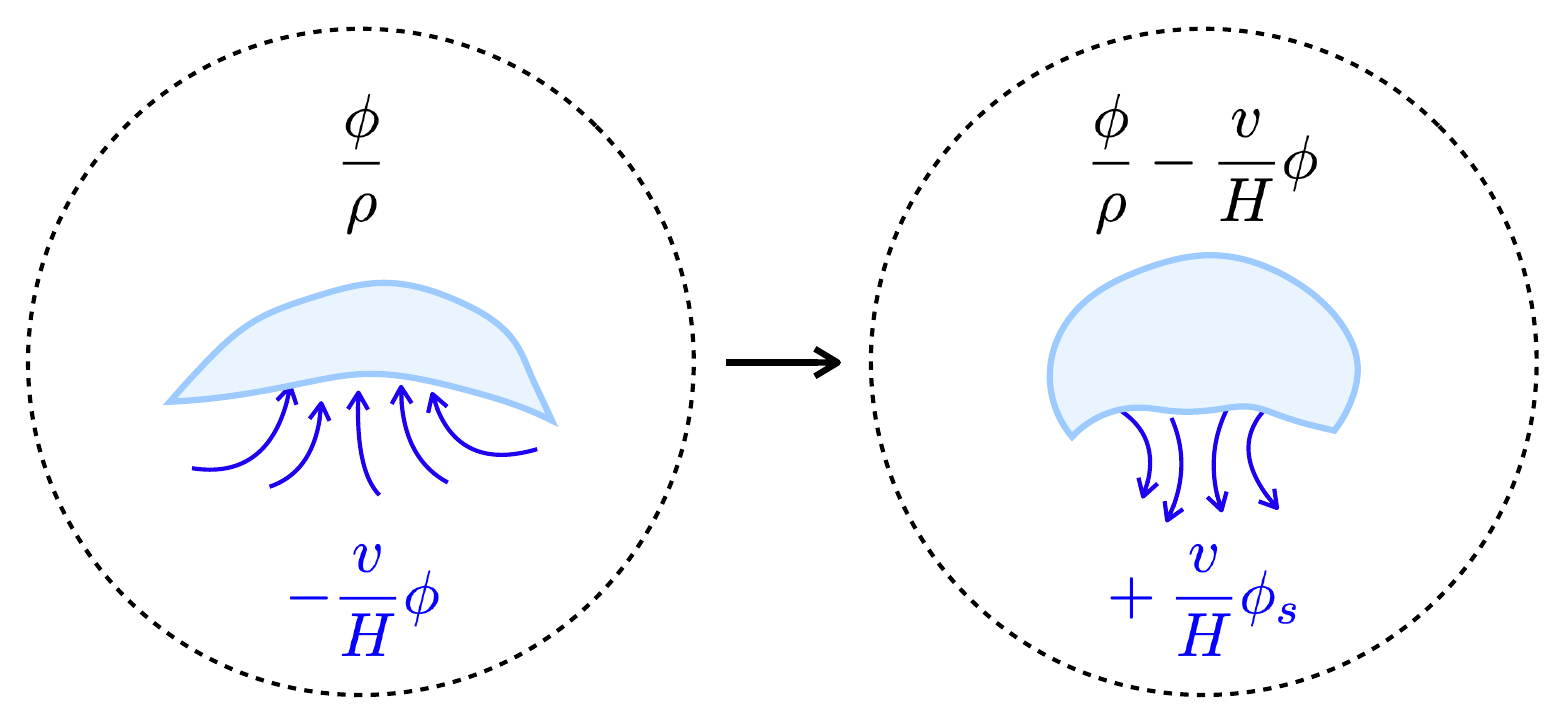}
\end{center}
\caption{\label{fig:release} Illustration of the chemical release mechanism. At the beginning of a bell cycle, a volume $v$ of water is drawn into the bell (left), enriched at concentration $\phi=\phi_s$ and released back in the water. Through this simplified mechanism, chemical concentration increases while remaining below the saturation point.}
\end{figure}

The model of \cite{gengel2024swarm} proposes chemotactic signalling as a possible mechanism to favor jellyfish aggregation. Such a mechanism is still unknown and currently under investigation, although previous studies have reported evidence that chemical cues influence (\textit{Aurelia}) jellyfish behaviour and habitat selection \cite{albert2011s,gimenez2025waterborne}. As explained in Section \ref{sec:agentbasedmodel}, signalling is modelled as a dependence of the steering vector field $\boldsymbol{\Psi}$ on the local gradient of concentration $\phi$ of a chemoattractant, which is assumed to be released by jellyfish. Chemical release is represented as a source term $S$ in the advection-diffusion equation \eqref{eq:signalling}. In \cite{gengel2024swarm}, this term couples a field, $\phi(\mathbf{x},t)$, with the discrete agents, which is numerically challenging. In contrast, our coarse-grained model allows us to express $S$ in a more straightforward manner, namely as a coupling between two fields, $\phi$ and $\rho$, which can be obtained by assuming the following scheme: during each bell cycle, jellyfish pump in a volume $v$ of surrounding water and then release the same volume of water enriched with chemoattractant, to saturation density defined as $\phi=\phi_s$. Now let us remind that, our model being two-dimensional due to the clear scale separation between the vertical (few metres) and horizontal (kilometres) dimensions of swarms, $\rho, \phi$ and $\phi_s$ must be understood as two-dimensional densities, whereas $v$ is actually a three-dimensional volume. For the sake of dimensional consistency, we thus introduce a vertical length $H$ representative of the swarm's vertical structure and jellyfish aspect ratio, such that two-dimensional densities are given by the product of the three-dimensional ones and $H$. Therefore, in the volume occupied by one jellyfish, which is $H/\rho$, the amount of chemical, initially of $\phi/\rho$, becomes $(1/\rho - v/H) \phi + \phi_s v/H$ after a bell cycle, which yields an increase $\rho (\phi_s - \phi)v/H$ of chemoattractant concentration over a time $2 \pi / \Omega$ (Figure \ref{fig:release}). An exponential sink term $-\phi/\tau$ is added as well to model natural decay effects such as sedimentation, vertical transport, or microbial consumption \cite{gengel2024swarm}. To summarise, we adopt the following source-sink:
\begin{equation} \label{eq:source_term}
    S = K (\phi_s - \phi) \rho - \frac{\phi}{\tau} \ ,
\end{equation}
where $K = v \Omega / (2 \pi H)$ is a coefficient that couples both two-dimensional densities $\rho$ and $\phi$. Consistently with \cite{gengel2024swarm}, chemical release increases proportionally with $\rho$, and slows down progressively as the saturation level ($\phi = \phi_s$) is approached. Once again, we emphasise the fact that this model, as well as the very existence of such chemical signalling mechanism, remain hypothetical to this day. Expression \eqref{eq:source_term} is intended to provide a workable theoretical framework. Further experimental data will be used to validate this model and refine parameters $K,\phi_s$ and $\tau$. Injecting Expression \eqref{eq:source_term} in \eqref{eq:signalling} yields the advection-diffusion equation
\begin{equation} \label{eq:signalling_source}
    \partial_t \phi + \boldsymbol{\nabla} \cdot \left( \phi \mathbf{U} \right) - D_\phi \boldsymbol{\nabla}^2 \phi = K (\phi_s - \phi) \rho - \frac{\phi}{\tau} \ .
\end{equation}

\subsection{Active steering mechanisms} \label{part:steering}

Finally, closing the system of equations requires to express the steering field $\boldsymbol{\Psi}$, i.e. listing the different stimuli and the corresponding reaction coefficients $\varepsilon_s$ of \eqref{eq:vectorpsi}. In \cite{gengel2024swarm} as well as this paper, we only consider positive rheotaxis (R), avoiding shear inducing turbulence (T), and chemical signalling (S) to be significant stimuli for the study of large swarms interacting with sea currents. Gengel \textit{et al} \cite{gengel2024swarm} adopt the following expressions:
\begin{equation} \label{eq:Erik_steering}
    \begin{split}
        \boldsymbol{\Psi} &= \varepsilon_\text{S} \frac{\boldsymbol{\nabla}\phi}{|\boldsymbol{\nabla}\phi|} - \varepsilon_\text{R} \frac{\mathbf{U}}{|\mathbf{U}|} - \varepsilon_\text{T} \frac{\boldsymbol{\nabla}\mathcal{C}}{|\boldsymbol{\nabla}\mathcal{C}|} \ ,\\[6pt]
        \text{with} \quad \varepsilon_\text{S} = \varepsilon_{\text{S},0} \mathcal{T}_\text{S} (&\phi) \ , \quad \varepsilon_\text{R} = \varepsilon_{\text{R},0} \mathcal{T}_\text{R} (|\mathbf{U}|) \ , \quad \text{and} \quad \varepsilon_\text{T} = \varepsilon_{\text{T},0} \mathcal{T}_\text{T} (\mathcal{C}) \ ,
    \end{split}
\end{equation}
where $\mathcal{C} = |\boldsymbol{\nabla} \times \mathbf{U}|$ is the absolute vorticity of the flow which is chosen to measure shear strength experienced by jellyfish, and $\varepsilon_{\text{S},0}, \varepsilon_{\text{R},0}$ and $\varepsilon_{\text{T},0}$ are constant parameters which characterise jellyfish reactivity to signalling, current and shear, respectively. Those reactivity times are thus assumed to be independent of the intensity of the stimuli themselves, except for the \textit{ignorance functions} $\mathcal{T}_\text{S}, \mathcal{T}_\text{R}$ and $\mathcal{T}_\text{T}$ which characterise the hypothetical absence of reaction of the jellyfish (see Expressions (9) and (10) in \cite{gengel2024swarm}) to, respectively, i/ the gradient of signalling chemical concentration -- if the concentration itself is too low or too high, close to saturation ($\mathcal{T}_\text{S}$ is zero under a certain threshold, sharply increases to $1$ beyond it, and then sharply vanishes again near the saturation point $\phi = \phi_s$),  ii/ the current -- if the latter is too weak ($\mathcal{T}_\text{R}$ is zero under a certain threshold and sharply increases to $1$ beyond this threshold) and iii/ similarly the gradient of absolute vorticity -- if shear is not strong enough ($\mathcal{T}_\text{T}$ is zero under a certain threshold and sharply increases to $1$ beyond it). In other words, the model adopted in \cite{gengel2024swarm} assumes that jellyfish respond to stimuli in an "on-off" manner, i.e. such that the response coefficients $\varepsilon_s$ do not have a continuous dependency on the intensity of the stimuli, but a stepwise one. Alternatively, one can assume a linear regime, i.e. that the current and the gradient of chemical are such that jellyfish response to those stimuli has a form
\begin{equation} \label{eq:steering_linear}
    \boldsymbol{\Psi} = \beta \phi \left( \phi_s - \phi \right) \boldsymbol{\nabla}\phi - \alpha \mathbf{U} - \gamma \mathcal{C} \boldsymbol{\nabla} \mathcal{C} \ ,
\end{equation}
which is the simplest expression translating a possible combined linear dependency of the steering field on:

\begin{itemize}
    \item $\phi$, $\phi_s - \phi$ and $\boldsymbol{\nabla}\phi$, meaning that reaction to signalling vanishes when the chemical concentration becomes small of close to the saturation point $\phi_s$ (inhibition), or when the concentration is spatially homogeneous.
    \item $\mathbf{U}$, meaning that positive rheotaxis is weak if the intensity of the current is weak.
    \item $\mathcal{C}$ and $\boldsymbol{\nabla}\mathcal{C}$, meaning that jellyfish avoid shear more effectively when the absolute vorticity and its gradient are larger.
\end{itemize}

At lowest order, one can assume that the coefficients $\alpha, \beta$ and $\gamma$ are constant (in theory, one should add a limitation to $\alpha$ and $\gamma$ for strong currents, but $|\mathbf{U}|$ generally does not exceed a few tens of cm s$^{-1}$, which is also comparable to the swimming speed of \textit{R. nomadica}). Whether one adopts a model like \eqref{eq:Erik_steering} or \eqref{eq:steering_linear}, the steering field is a superposition of the different directional cues, weighted by their relative importance. Just like for the rest of this model's parameters and variables, experimental data will be necessary to infer the validity of this model of steering field, and explore the details of jellyfish' response to different stimuli. 

\section{Hydrodynamic equations} \label{sec:hydrodynamics}

\subsection{Density, polarisation and continuity equation} \label{part:polarisation}

Naturally, the observable that one wants to track in swarms is not the distribution $f$ itself but rather the density $\rho$, for which a continuity equation can be obtained by integrating Equation \eqref{eq:Fokker-Planck} in $\theta$, using definition \eqref{eq:density}:
\begin{equation} \label{eq:continuity}
    \partial_t \rho + \boldsymbol{\nabla} \cdot \left( \rho \mathbf{U} + \mathcal{V} \mathbf{p} \right) = 0 \ ,
\end{equation}
where the polarisation $\mathbf{p}$ is defined as
\begin{equation}
    \mathbf{p} (\mathbf{x},t) \equiv \rho \langle \mathbf{e}_\theta \rangle \equiv \int_{-\pi}^{\pi} f (\mathbf{x},\theta,t) \mathbf{e}_\theta \, \dd \theta \ .
\end{equation}

Let us note that all angular terms -- i.e. the RHS of \eqref{eq:Fokker-Planck} -- yield no contribution to the continuity equation \eqref{eq:continuity}, since they are expressed as derivatives in $\theta$. However, this is a more general result which would be obtained as well, say, for a different collision operator $I_\text{col}[f,f]$, because the angular processes shown in Figure \ref{fig:process} affect jellyfish orientation, but not their number. From this point on, it will be convenient to define the \textit{angular modes} of the distribution $f$, i.e. its Fourier coefficients:
\begin{equation} \label{eq:Fourier_moments}
    f_n (\mathbf{x},t) \equiv \int_{-\pi}^{\pi} f (\mathbf{x},\theta,t) \ee^{\ii n \theta} \, \dd \theta \ ,
\end{equation}
which quantify the weight of each angular order. In particular, $f_0 = \rho$ and $f_1$ is the complex-plane representation of the polarisation $\mathbf{p}$ \cite{bertin2006boltzmann}. An equation for $\partial_t  \mathbf{p}$ can be obtained as well by integrating Equation \eqref{eq:Fokker-Planck} multiplied by $\mathbf{e}_\theta$, which involves the next-order mode $f_2$ (related to the \textit{nematic tensor} \cite{marchetti2013hydrodynamics}), because of the active transport (self-propulsion) term $\boldsymbol{\nabla} \cdot ( f \mathcal{V} \mathbf{e}_\theta )$ and the active steering term $- \partial_\theta \left( f \Psi \sin \left( \theta - \theta_\Psi \right) \right)$ (see the projected Equation \eqref{eq:Fourier_appendix} in appendix B), for which an evolution equation involving $f_3$ must be solved, and so on. Closure is usually achieved by cutting off all higher-order modes above a certain degree. For instance, with the Vicsek model, keeping terms up to the second order ($n=2$) yields the Toner-Tu equations \cite{toner1995long,bertin2006boltzmann}, which is the minimal form that exhibits a phase transition leading to spontaneous flocking. Such approximate hydrodynamic closure is valid only if the modes decrease with a power law as $f_n \sim \rho r^n$, with $r \ll 1$. In the present case, where dipolar steering is a significant factor, a similar hydrodynamic closure is possible if the \textit{alignment parameter}
\begin{equation}
    \kappa (\mathbf{x},t) = \frac{\Psi}{\mathcal{D}(\rho)}
\end{equation}
is small, i.e. in regions where stimuli-induced steering is much slower than angular diffusion due to stochastic self-diffusion and/or collisions. This limit is treated in appendix B and yields an extension of the Keller-Segel model \cite{keller1971model}. However, jellyfish are believed to respond fast to stimuli, when those are strong enough. Typically, in \cite{gengel2024swarm}, $\Psi$ can be of order $0.1$ rad s$^{-1}$ while $D_r \approx 4 \ 10^{-3}$ rad$^2$ s$^{-1}$, although $\Psi$ can also become very small if stimuli are too weak, which can be characterised for instance by the ignorance functions \eqref{eq:Erik_steering}. In other words, both limits can coexist, but if $\kappa$ is not generally a small parameter, one must adopt another point of view, which relies on the separation between individual and collective time scales.

\subsection{Hydrodynamic closure in the limit of fast angular relaxation} \label{part:closure}

The dynamics described by Equation \eqref{eq:Fokker-Planck} involves a strong separation between the time scales of individual dynamics and that of collective swarm evolution. The latter is set by the advection of the aggregation by the ambient flow, characterised by a time scale $T_\text{adv} = L / U_0$, where $U_0$ is a characteristic current velocity and $L$ the characteristic spatial scale of the swarms that we aim to describe. Oceanographic observations in the Eastern Mediterranean report ambient-current velocities ranging from approximately $10^{-2}$ to $10^{-1}$ m s$^{-1}$ \cite{feliks2022intraseasonal,mantel2024seasonal}, while \textit{R. nomadica} aggregations accomodated by these currents can extend over spatial scales of many kilometres \cite{edelist2022tracking,douek2024long}. Moreover, direct measurements reveal that \textit{R. nomadica} swimming speed is comparable to the velocity of the currents transporting them \cite{malul2019levantine}, so that we can consider that $\mathcal{V}$ is also of order $U_0$. Therefore, the corresponding advective time scale is typically of order days or longer. In contrast, the characteristic time scales associated with individual orientation dynamics -- rotational diffusion, active steering, and collisions -- are of order seconds to minutes \cite{albert2011s}, and thus act on much faster time scales. We therefore distinguish the fast individual dynamics from the slow evolution of the swarm. In other words, we assume
\begin{equation}
    \mathcal{D} T_\text{adv} \gg 1 \quad \text{and} \quad \Psi T_\text{adv} \gg 1 \ ,
\end{equation}
and we introduce the dimensionless parameter
\begin{equation}
    \epsilon \equiv \frac{1}{D_r T_\text{adv}} = \frac{U_0}{L D_r} \ll 1 \ .
\end{equation}

Importantly, this separation does not depend on whether the steering or rotational diffusion dominates the angular dynamics. The dimensionless ratio $\kappa = \Psi / \mathcal{D}$ controls the relative importance of these two mechanisms, whereas both remain fast compared with the advective dynamics as long as $1/\epsilon, \kappa / \epsilon \gg 1$. Up to $f$, the terms of the LHS of Equation \eqref{eq:Fokker-Planck} -- which characterise temporal and spatial advection -- are of order $1/T_\text{adv} = \epsilon D_r$, whereas the terms of the RHS -- which embody the fast angular dynamics -- are of order $\Psi$ and $\mathcal{D}$. Consequently, the angular distribution rapidly relaxes compared with the evolution of the swarm density. The leading-order angular structure is therefore determined by the stationary angular equation
\begin{equation}
    \mathcal{L}_\theta [f] = 0 \ ,
\end{equation}
whose $2 \pi$-periodic solution provides the local quasi-equilibrium angular distribution:
\begin{equation} \label{eq:Bessel_solution}
    \begin{split}
        f^\text{vM} (\mathbf{x},\theta,t) &= \frac{\rho}{2 \pi \mathcal{I}_0 (\kappa)} \exp \left( \boldsymbol{\kappa} \cdot \mathbf{e}_\theta \right) \ ,\\[6pt]
        \text{with} \quad \boldsymbol{\kappa} = \frac{\boldsymbol{\Psi}}{\mathcal{D}(\rho)} = \kappa \mathbf{e}_{\theta_\Psi} \quad &\text{and} \quad \mathcal{I}_n (z) = \frac{1}{2 \pi} \int_{-\pi}^{\pi} \ee^{\ii n \theta} \ee^{z \cos \theta} \, \dd \theta \ .
    \end{split}
\end{equation}

\begin{figure}[h]
\begin{center}
    \includegraphics[scale=0.43]{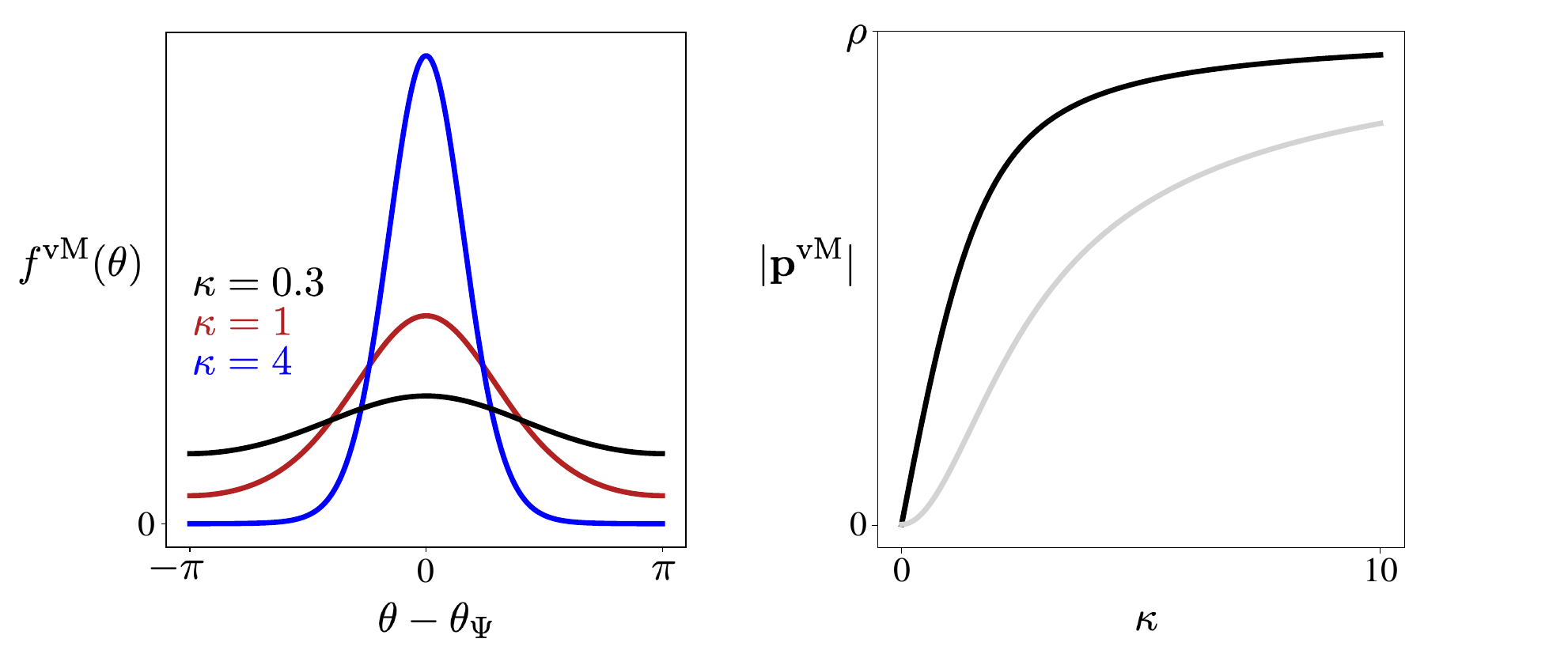}
\end{center}
\caption{\label{fig:vonMises_polarisation} The von Mises angular distribution. Left: $f^\text{vM}(\theta)$ for different values of the alignment parameter. The larger $\kappa$, the sharper is the distribution around the steering angle $\theta_\Psi$. Right: Norm of the polarisation $\mathbf{p}^\text{vM}$ as a function of $\kappa$ (black). The quantity $\rho \mathcal{I}_2 (\kappa) / \mathcal{I}_0 (\kappa)$ is also shown in light gray.}
\end{figure}

The functions $\mathcal{I}_n$ are known as the modified Bessel functions of the first kind, and the distribution $f^\text{vM}$ of Expression \eqref{eq:Bessel_solution} is the angular von Mises distribution \cite{von1918ganzzahligkeit,silvano2026hydrodynamic}. The alignment parameter $\kappa$, which is a function of time $t$ and position $\mathbf{x}$ (as well as the density $\rho$), measures how peaked the angular distribution is, locally (Figure \ref{fig:vonMises_polarisation}). By definition, the angular modes of this distribution \eqref{eq:Bessel_solution} are
\begin{equation}
    f_n^\text{vM} = \rho \frac{\mathcal{I}_n (\kappa)}{\mathcal{I}_0 (\kappa)} \ee^{i n \theta_\Psi} \ .
\end{equation}

In particular for $n=1$, the polarisation $\mathbf{p} = \rho \langle \mathbf{e}_\theta \rangle$ is approximately given by
\begin{equation} \label{eq:p_vM}
    \mathbf{p}^\text{vM} = \rho \frac{\mathcal{I}_1 (\kappa)}{\mathcal{I}_0 (\kappa)} \mathbf{e}_{\theta_\Psi} \ .
\end{equation}

Therefore, $\kappa$ appears as a dimensionless parameter that characterises the local directionality of collective swimming, since $\Psi$ measures the reactivity of jellyfish to environmental stimuli whereas $\mathcal{D}$ measures disorientation, as a result of self-diffusion ($D_r$) and direct collisions ($\delta$). According to Expression \eqref{eq:Bessel_solution}, the angular distribution is thus very peaked around the steering angle $\theta_\Psi$ for large values of $\kappa$, and the bandwidth is a manifestation of local disorientation (Figure \ref{fig:vonMises_polarisation}, left). Accordingly, the order parameter $|\langle \mathbf{e}_\theta \rangle_\text{vM}| = \mathcal{I}_1 (\kappa) / \mathcal{I}_0 (\kappa)$, which measures collective alignment locally, increases with $\kappa$ and converges towards $1$ for large $\kappa$ (Figure \ref{fig:vonMises_polarisation}, right). For small values of $\kappa$, the Taylor expansion of $\mathcal{I}_n (\kappa)$ \cite{abramowitz1948handbook} reads as
\begin{equation} \label{eq:Taylor_vM}
    f_n^\text{vM} (\mathbf{x},t) \sim \frac{\rho}{n!} \left( \frac{\kappa}{2} \right)^n \ee^{\ii n \theta_\Psi} \quad \text{for} \quad \kappa \ll 1 \ ,
\end{equation}
which is consistent with the scaling $f_n \sim \rho \kappa^n$ used in \cite{bertin2006boltzmann} and mentioned earlier in this subsection (see appendix B). In particular, \eqref{eq:Taylor_vM} yields
\begin{equation} \label{eq:pvm_limit}
    \mathbf{p}^\text{vM} \sim \rho \frac{\boldsymbol{\kappa}}{2} = \rho \frac{\boldsymbol{\Psi}}{2 \mathcal{D}} \quad \text{for} \quad \kappa \ll 1 \ .
\end{equation}

Conversely, for large values of $\kappa$, $f_n \sim \rho$ for all $n$, therefore all modes $f_n$ equally matter in this limit. As discussed in \ref{part:polarisation}, $\kappa$ can range from small to large values in the same region, thus a unified hydrodynamic equation must account for both limits. This is where the separation of time scales allows us to establish a closed hydrodynamic equation which is valid for all values of $\kappa$. Indeed, as we will demonstrate explicitly in the next subsection, we have
\begin{equation} \label{eq:continuity_leading_order}
    \partial_t \rho + \boldsymbol{\nabla} \cdot \left( \rho \mathbf{U} + \rho \mathcal{V} \frac{\mathcal{I}_1 (\kappa)}{\mathcal{I}_0 (\kappa)} \mathbf{e}_{\theta_\Psi} \right) = \mathcal{O} \left( \rho D_r \epsilon^2 \right) \ ,
\end{equation}
while the terms of the LHS of \eqref{eq:continuity_leading_order} are of order $\rho D_r \epsilon$. Expression \eqref{eq:continuity_leading_order} contains two transport terms between brackets: the first one accounts for passive transport by the flow, whereas the second characterises collective directional swimming. For $\epsilon \ll 1$, one can thus take the RHS of \eqref{eq:continuity_leading_order} to be zero -- the angular adjustment to the von Mises distribution is therefore regarded as an instantaneous process -- and directly use the hyperbolic (i.e. purely advective) continuity equation. The set of coupled equations

\begin{equation}
    \boxed{\begin{split}
        &\partial_t \rho + \boldsymbol{\nabla} \cdot \left( \rho \mathbf{U} + \rho \mathcal{V} \frac{\mathcal{I}_1 (\kappa)}{\mathcal{I}_0 (\kappa)} \mathbf{e}_{\theta_\Psi} \right) = 0 \ ,\\[6pt]
        &\kappa \mathbf{e}_{\theta_\Psi} \equiv \frac{\boldsymbol{\Psi}}{\mathcal{D}(\rho)} \equiv \frac{1}{\mathcal{D}(\rho)} \left( \beta \phi \left( \phi_s - \phi \right) \boldsymbol{\nabla}\phi - \alpha \mathbf{U} - \gamma \mathcal{C} \boldsymbol{\nabla} \mathcal{C} \right) \ ,\\[6pt]
        &\partial_t \phi + \boldsymbol{\nabla} \cdot \left( \phi \mathbf{U} \right) - D_\phi \boldsymbol{\nabla}^2 \phi = K (\phi_s - \phi) \rho - \frac{\phi}{\tau} \ ,
    \end{split}}
\end{equation}
along with initial and boundary conditions -- and the definitions of the variables and parameters contained in those equations --, constitutes a closed system that can be readily implemented in an ocean-current model -- such as, for instance, the \texttt{MITgcm} \cite{marshall1997finite} -- to simulate the dynamics of large swarms.

\subsection{Next order contributions to the hydrodynamic closure}

While the analysis in the previous subsection relates to large swarms that are relatively dilute, one may extend the hydrodynamic closure to describe dense structures at smaller scales (see for instance the patches of aggregation visible in Figure 1 of \cite{kuplik2026rhopilema}), for which $\epsilon$ remains small but may take rather moderate values. This implies including some of the next-order terms contained in the RHS of Equation \eqref{eq:continuity_leading_order}. To infer those corrections, we write the distribution $f$ as the sum of the dominant von Mises distribution \eqref{eq:Bessel_solution} and a small perturbation $\epsilon f^\epsilon$ (of order $\epsilon \rho$):
\begin{equation}
    f = f^\text{vM} + \epsilon f^\epsilon \ .
\end{equation}

By definition of the density $\rho$ and $f^\text{vM}$, $f^\epsilon$ has the following properties:
\begin{equation}
    \begin{split}
        f^\epsilon_0 &= \int_{-\pi}^{\pi} f^\epsilon(\mathbf{x},\theta,t) \, \dd \theta = 0 \ ,\\
        f^\epsilon_1 &= \int_{-\pi}^{\pi} f^\epsilon(\mathbf{x},\theta,t) \mathbf{e}_\theta \, \dd \theta = \mathbf{p}^\epsilon = \frac{\mathbf{p} - \mathbf{p}^\text{vM}}{\epsilon} \ .
    \end{split}
\end{equation}

Injecting these definitions into Equation \eqref{eq:Fokker-Planck} yields
\begin{equation} \label{eq:Boltzmann_perturbation}
    \partial_t f + \boldsymbol{\nabla} \cdot \left( f \left( \mathbf{U} + \mathcal{V} \mathbf{e}_\theta \right) \right) = \epsilon \mathcal{L}_\theta \left[ f^\epsilon \right] \ ,
\end{equation}
where $\mathcal{L}_\theta$ is the fast angular operator defined in Equation \eqref{eq:Fokker-Planck}, whose kernel contains $f^\text{vM}$. In angular Fourier space, Equation \eqref{eq:Boltzmann_perturbation} reads as (see appendix B)
\begin{equation} \label{eq:Boltzmann_Fourier}
    \partial_t f_n + \boldsymbol{\nabla} \cdot \left( f_n \mathbf{U} \right) + \frac{1}{2} \underline{\nabla} \left( f_{n-1} \mathcal{V} \right) + \frac{1}{2} \underline{\nabla}^\star \left( f_{n+1} \mathcal{V} \right) = \epsilon \left( -n^2 \mathcal{D}(\rho) f^\epsilon_n + \frac{n}{2} \underline{\Psi} f^\epsilon_{n-1} - \frac{n}{2}
    \underline{\Psi}^\star f^\epsilon_{n+1} \right) \ ,
\end{equation}
for all $n \in \mathbb{Z}$, where we also define the complex field $\underline{\Psi} = \Psi_x + \ii \Psi_y$ and operator $\underline{\nabla} = \partial_x + \ii \partial_y$, $^\star$ being the complex conjugation. By definition \eqref{eq:Fourier_moments}, for all $n$, we have $f_{-n} = f_n^\star$. The case $n=0$, for which the RHS of \eqref{eq:Boltzmann_Fourier} vanishes, yields the continuity equation \eqref{eq:continuity} already introduced at the beginning of this section, which now takes the following form:
\begin{equation} \label{eq:continuity_unified}
    \partial_t \rho + \boldsymbol{\nabla} \cdot \left( \rho \mathbf{U} + \rho \mathcal{V} \frac{\mathcal{I}_1 (\kappa)}{\mathcal{I}_0 (\kappa)} \mathbf{e}_{\theta_\Psi} + \mathcal{V} \epsilon \mathbf{p}^\epsilon \right) = 0 \ .
\end{equation}

As for the remaining orders ($n \neq 0$), we can neglect the transport terms -- i.e. the LHS of \eqref{eq:Boltzmann_Fourier} -- applied to $f^\epsilon$, which are of order $\rho D_r \epsilon^2$ whereas the other terms are of order $\rho D_r \epsilon$. We thus get an approximate recurrence relation for the functions $f_n^\epsilon$:
\begin{equation} \label{eq:Boltzmann_Fourier_recurrence}
    \begin{split}
        &\epsilon \left( -n^2 \mathcal{D}(\rho) f^\epsilon_n + \frac{n}{2} \underline{\Psi} f^\epsilon_{n-1} - \frac{n}{2} \underline{\Psi}^\star f^\epsilon_{n+1} \right)\\[6pt]
        \approx \: &\partial_t f^\text{vM}_n + \boldsymbol{\nabla} \cdot \left( f^\text{vM}_n \mathbf{U} \right) + \frac{1}{2} \underline{\nabla} \left( f^\text{vM}_{n-1} \mathcal{V} \right) + \frac{1}{2} \underline{\nabla}^\star \left( f^\text{vM}_{n+1} \mathcal{V} \right) \ .
    \end{split}
\end{equation}

Considering only the first harmonic, $f^\epsilon_1$ (dipolar correction), and neglecting the others, Equation \eqref{eq:Boltzmann_Fourier_recurrence} yields
\begin{equation} \label{eq:Boltzmann_firs-order-perturbation}
    -\mathcal{D(\rho)} \epsilon f^\epsilon_1 \approx \partial_t \left( \rho \frac{\mathcal{I}_1 (\kappa)}{\mathcal{I}_0 (\kappa)} \ee^{i \theta_\Psi} \right) + \boldsymbol{\nabla} \cdot \left( \rho \frac{\mathcal{I}_1 (\kappa)}{\mathcal{I}_0 (\kappa)} \ee^{i \theta_\Psi} \mathbf{U} \right) + \frac{1}{2} \underline{\nabla} \left( \mathcal{V} \rho \right) + \frac{1}{2} \underline{\nabla}^\star \left( \mathcal{V} \rho \frac{\mathcal{I}_2 (\kappa)}{\mathcal{I}_0 (\kappa)} \ee^{2 i \theta_\Psi} \right) \ ,
\end{equation}
which can be expressed back in terms of real-valued vector fields as
\begin{equation} \label{eq:complex_diffusion}
    \begin{split}
        -\mathcal{D(\rho)} \epsilon \mathbf{p}^\epsilon &\approx \partial_t \mathbf{p}^\text{vM} + \left( \mathbf{U} \cdot \boldsymbol{\nabla} \right) \mathbf{p}^\text{vM} + \left( \boldsymbol{\nabla} \cdot \mathbf{U} \right) \mathbf{p}^\text{vM} + \frac{1}{2} \boldsymbol{\nabla} \left( \mathcal{V} \rho \right)\\[6pt]
        &+ \left( \boldsymbol{\nabla} \cdot \left( G \boldsymbol{\kappa} \right) \right) \boldsymbol{\kappa} + \left( G \boldsymbol{\kappa} \cdot \boldsymbol{\nabla} \right) \boldsymbol{\kappa} - \frac{1}{2} \boldsymbol{\nabla} \left( G \boldsymbol{\kappa}^2 \right) \ ,
    \end{split}
\end{equation}
where we have introduced the quantity
\begin{equation} \label{eq:G}
    G = \frac{\mathcal{V} \rho \mathcal{I}_2 (\kappa)}{\kappa^2 \mathcal{I}_0 (\kappa)} \ ,
\end{equation}
which has a finite limit, $\rho \mathcal{V}/8$, when $\kappa \ll 1$, in virtue of the expansion \eqref{eq:Taylor_vM}. Expression \eqref{eq:complex_diffusion} is a complex term that combines classical diffusion (fourth term) with other contributions accounting for the transport of mean polarisation (first three terms) and nematic tensor (second line), which come from the last term of Equation \eqref{eq:Boltzmann_firs-order-perturbation}. In the quasi-isotropic regime $\kappa \ll 1$ (see appendix B), classical diffusion dominates and the coarse-grained flux reads as
\begin{equation} \label{eq:KS_flux}
    \mathcal{V} \mathbf{p} \approx \frac{\mathcal{V}}{2 \mathcal{D}} \left( \rho \boldsymbol{\Psi} - \boldsymbol{\nabla} \left( \mathcal{V} \rho \right) \right) \ .
\end{equation}

In other words, the mean active transport is proportional to $\boldsymbol{\Psi}$, with a macroscopic diffusivity of $\mathcal{V}^2 / 2 \mathcal{D}$, which naturally increases with mobility ($\mathcal{V}$) and decreases with angular disorder and collisions ($\mathcal{D}$). To the extent that the steering field $\boldsymbol{\Psi}$ is partly aligned with the gradient of signalling tracer's concentration $\boldsymbol{\nabla} \phi$, Equation \eqref{eq:continuity} with \eqref{eq:KS_flux} is strongly similar to the continuity equation of the Keller-Segel model (with additional passive transport), in which the coarse-grained active velocity consists of a diffusive contribution and a term proportional to the gradient of chemical concentration \cite{keller1971model}. In the general case where $\kappa$ can be arbitrarily large, all terms in Expression \eqref{eq:complex_diffusion} have contributions of the same order, yielding a complex correction which differs from the classical diffusion term flux $-\mathcal{V} \boldsymbol{\nabla} \left( \mathcal{V} \rho \right) / 2 \mathcal{D}$. For instance, combining the fourth and last terms of the RHS of \eqref{eq:complex_diffusion} yields the contribution
\begin{equation}
    -\frac{\mathcal{V}}{2 \mathcal{D}} \boldsymbol{\nabla} \left( \mathcal{V} \rho - G \kappa^2 \right) = -\frac{\mathcal{V}}{2 \mathcal{D}} \boldsymbol{\nabla} \left( \left( 1 - \frac{\mathcal{I}_2 (\kappa)}{\mathcal{I}_0 (\kappa)} \right) \mathcal{V} \rho \right)
\end{equation}
for the diffusion term, which vanishes as $\kappa$ becomes large, since $\mathcal{I}_2 (\kappa) / \mathcal{I}_0 (\kappa) \rightarrow 1$ as $\kappa \rightarrow + \infty$ (Figure \ref{fig:vonMises_polarisation}). In all cases, all terms of \eqref{eq:complex_diffusion} are dominated by $\rho D_r \epsilon^2$ regardless of the value of $\kappa$, which means that this corrective flux remains small for large-scale phenomena, although it should be taken into account especially at the boundaries of jellyfish aggregations, where sharp density fronts may form. However, if the continuity equation is integrated using finite-volume methods, those numerical schemes are known to produce artificial diffusive effects, with diffusivity $\sim \mathcal{V} \Delta x/2$ (where $\Delta x$ is the grid step size), whereas physical diffusivity is of order $\mathcal{V}^2 / 2 \mathcal{D}$. In other words, physical diffusivity is numerically irrelevant if the spatial step $\Delta x$ is larger than the mean free path $\mathcal{V} / \mathcal{D}$, which is roughly a few tens of metres at best, and decreases in dense aggregations.

\subsection{Roadmap towards ocean-current models and forecasting} \label{part:roadmap}

One can implement the continuity equation \eqref{eq:continuity_unified} -- either with $\mathbf{p}^\epsilon = 0$ (purely advective or hyperbolic form), or using Expression \eqref{eq:complex_diffusion} -- with properly defined initial and boundary conditions on the domain, using a prescribed current $\mathbf{U}$ extracted from ocean-current simulations or available data. Besides the density $\rho$ which is the field to be computed, the other necessary variables to be defined in the continuity equation are $\mathcal{V}$, $\mathcal{D}(\rho), \boldsymbol{\kappa}, \mathbf{p}^\text{vM}$ and $G$. The last four are defined by Expressions \eqref{eq:D}, \eqref{eq:Bessel_solution}, \eqref{eq:p_vM} and \eqref{eq:G}, respectively, and $\boldsymbol{\kappa}$ involves the flow velocity, its vorticity and the gradient of chemical concentration, as explained in \ref{part:steering}. The system of PDE is thus closed by the advection-diffusion equation \eqref{eq:signalling_source}, which is coupled to the continuity equation \eqref{eq:continuity_unified}. This coupling manifests in Equation \eqref{eq:continuity_unified} through the alignment field $\boldsymbol{\kappa}$, while it does through the density $\rho$ in the RHS of \eqref{eq:signalling_source}.\\

\begin{figure}[h]
\begin{center}
    \includegraphics[scale=0.53]{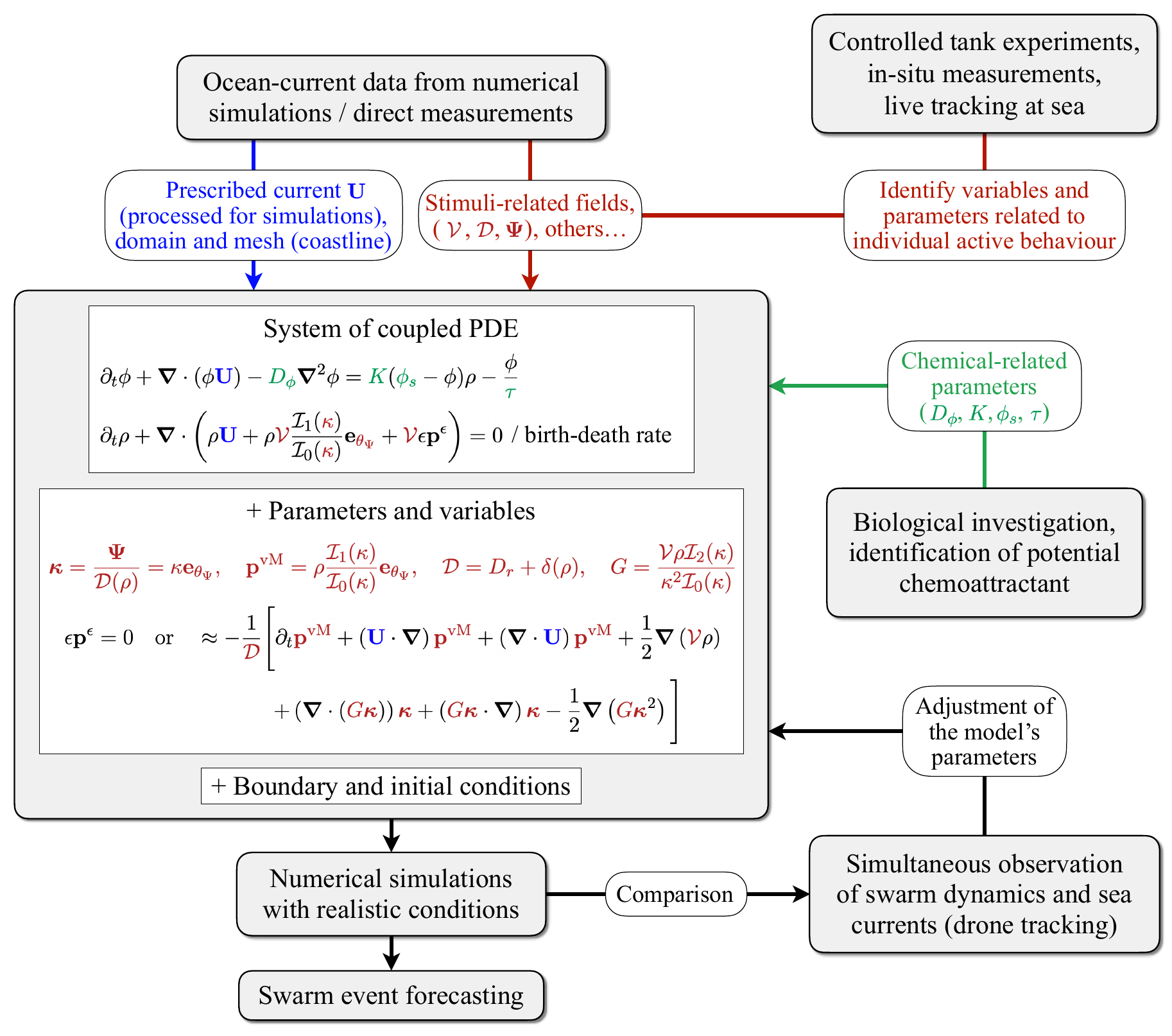}
\end{center}
\caption{\label{fig:roadmap} A roadmap to test and improve the model presented in this paper. This protocol bridges theory, numerical simulations embedding active-matter equations in ocean-current models, experiments and data analysis, the goal being to make this model accurate towards swarm events forecasting.}
\end{figure}

Figure \ref{fig:roadmap} illustrates the workflow which is being implemented in order to benchmark this model and improve it, bridging the theory that was introduced in \cite{gengel2023physics,gengel2024swarm}, as well as the present paper, with numerical simulations and experimental studies, including

\begin{itemize}
    \item \textbf{Biology and chemistry}: the existence of a signalling mechanism -- and a potential chemical for it -- is currently under investigation. Besides, studies have identified other factors that were not mentioned in this paper, which evidently affect certain aspects of jellyfish dynamics. For instance, the early development of jellyfish (including strobilation) is very sensitive to temperature \cite{dror2024rising,dror2025substrate}. This could be accounted for by adding a temperature-dependent source term in the RHS of Equation \eqref{eq:continuity_unified}, characterising birth-death rate under local conditions. Vodopivec and Malej \cite{vodopivec2025spatially} also focus on this aspect to model the long-term dynamics of jellyfish population in the Mediterranean Sea, including its dependence to food availability.
    \item \textbf{Tank-controlled and in-situ measurements}: the most empirical assumption of this paper is the existence of a field $\boldsymbol{\Psi}$ controlling an ad hoc dipolar self-steering of jellyfish in response to directional stimuli, whose hypothetical expressions are proposed in \ref{part:steering}. This model must be assessed with experiments in which jellyfish are exposed to controlled stimuli and one can measure their individual motion in response to those. In particular, one can simultaneously measure jellyfish motion and the velocity of the surrounding flow, either by using PTV \cite{shnapp2022myptv} in a lab environment, or by tracking jellyfish at sea using electronic tags \cite{diamant2023remote,sauviat2025jellyfish,gemmell2025movement}, for instance. Such experiments could, in principle, allow one to infer the parameters -- and their dependence on external variables -- characterising active swimming, i.e. the amplitude $\mathcal{V}$, the angular diffusivity $D_r$ and the steering field $\boldsymbol{\Psi}$. Additional stimuli must be tested, such as food (foraging), temperature, salinity, etc.
    \item \textbf{Swarm tracking and monitoring}: in the past few years, field campaigns have been carried out in the eastern Mediterranean Sea to track jellyfish swarms and study their collective swimming patterns using drones and identification tools \cite{malul2024directional}. Such campaigns could provide valuable and necessary information regarding the collective behaviour of jellyfish under realistic marine conditions (Malul \textit{et al} \cite{malul2024directional} suggest for instance that \textit{R. nomadica} jellyfish tend to collectively swim against the direction of surface gravity waves), and quantitative feedback to be analysed in light of the results of numerical simulations, in order to adjust the parameters of the model according to reality.
\end{itemize}

\section{Discussion and conclusion} \label{sec:conclusion}

In this paper, we have developed a continuous kinetic theory to describe the large-scale dynamics of jellyfish swarms, starting from an agent-based description of individual behaviour. We further showed that, as orientational dynamics involves time scales which are much shorter than those associated with spatial transport, this kinetic description admits a hydrodynamic closure based on the local angular distribution. This provides reduced equations for the jellyfish density, connecting individual behavioural mechanisms to population-scale dynamics through a set of coupled equations that can ultimately be incorporated into realistic ocean-circulation models.\\

An important question that this framework makes accessible is the relative contribution of passive transport by sea currents and active individual motion to the formation and persistence of observed jellyfish aggregations. Ocean currents alone can concentrate passive tracers through convergence and other features of the flow, and such passive mechanisms are likely to contribute to the formation of biological aggregations. Jellyfish, however, can actively swim relative to the surrounding currents and respond to environmental cues, potentially allowing them to reinforce, oppose or escape such passive accumulation. Their ability to maintain directed motion in weak currents may also contribute to the persistence of coherent aggregations. The present study provides a means of investigating how these effects combine in heterogeneous oceanic environments, e.g. fronts, shear zones and wave-driven flows. More generally, it allows competing behavioural mechanisms, such as directional responses to the surrounding flow and aggregation through chemical signalling, to be considered within a common description. Determining how the interplay between those mechanisms gives rise to persistent, transient or irregular aggregation patterns, and whether they can produce collective instabilities or distinct dynamical regimes, represents an important direction for future work.\\

A crucial next step is to implement the reduced equations in numerical ocean-current models and confront their predictions with observations. The roadmap developed here provides a basis for combining remote observations, in-situ measurements and controlled experiments to constrain the parameters of the model. Such comparisons will be essential for determining which of the proposed mechanisms are relevant to different jellyfish species and environmental conditions, and for assessing whether the resulting model can reproduce the spatial and temporal characteristics of observed swarms. The same framework can subsequently be extended to account for three-dimensional motion, more detailed collision dynamics and additional biological processes such as foraging and population turnover. These developments should also clarify which aspects of the two-dimensional description are sufficient at the scale of interest and which require a fully three-dimensional treatment.\\

More broadly, the framework developed here provides a route for bridging individual behaviour and ocean-scale ecological dynamics. Rather than treating large jellyfish aggregations solely as passive tracers of ocean circulation, it provides a means of incorporating active swimming, sensory responses and interactions into a continuum description that remains compatible with large-scale environmental models. This connection between behavioural mechanisms and population-scale transport may ultimately help determine when and where massive jellyfish swarms emerge, and provide a theoretical foundation for their numerical prediction in realistic ocean flows.\\

\noindent {\large \textbf{Data availability}}

\noindent No datasets were generated or analysed during this study.\\

\noindent {\large \textbf{Code availability}}

\noindent No code was used in this study.\\

\noindent {\large \textbf{Acknowledgements}}

\noindent The authors wish to thank Hila Dror, Ran Eisenberg, Ron Shnapp and Dror Angel for very useful discussions. N.P. is funded by the Azrieli Fellows Program, and by Tel Aviv University’s postdoctoral program. E.G. thanks the Minerva Stiftungsgesellschaft fuer die Forschung mbH for its support. This study is supported by the Israeli Science Foundation (grant No. 1218/23).\\

\noindent {\large \textbf{Author contributions}}

\noindent N.P. wrote the article. N.P., E.G. and E.H. developed the theoretical framework and performed the analysis. All authors conceived the study, discussed the results and contributed to the final manuscript.\\

\noindent {\large \textbf{Competing interests}}

\noindent The authors declare no competing interests.

\section*{Appendix A: Reduction of the collision operator}

\begin{figure}[h]
\begin{center}
    \includegraphics[scale=0.5]{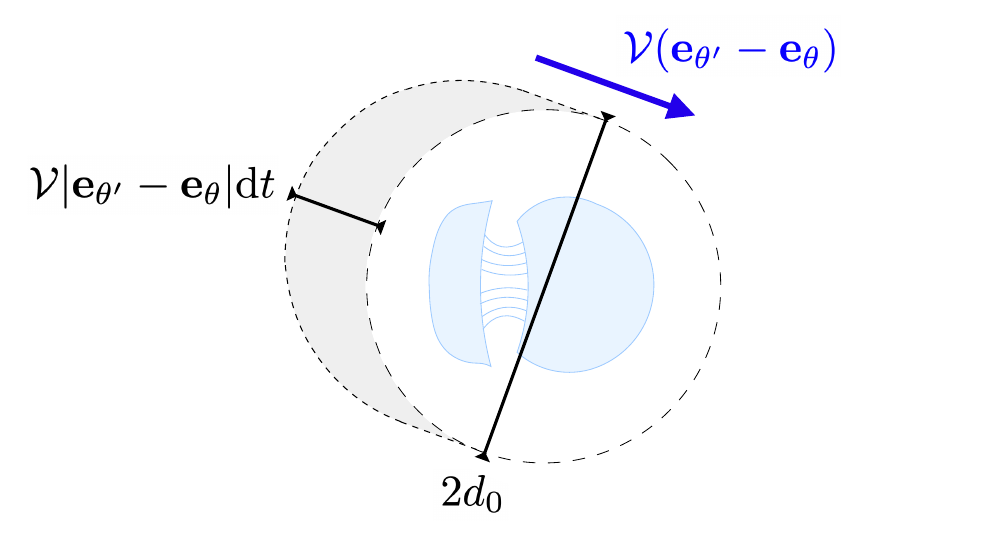}
\end{center}
\caption{\label{fig:collision} Counting the number of jellyfish of orientation $\theta'$ that collide with one of orientation $\theta$ during the time interval $\dd t$. A jellyfish of orientation $\theta$ is represented in the centre, and the motion of surrounding agents with orientation $\theta'$ is considered in its moving reference frame. They thus move with velocity $\mathcal{V} \left( \mathbf{e}_{\theta'} - \mathbf{e}_{\theta} \right)$, therefore pairwise collision is likely to occur with agents in the gray area, whose surface is given by $2 d_0 \mathcal{V} | \mathbf{e}_{\theta'} - \mathbf{e}_{\theta} | \dd t$, hence the expression \eqref{eq:collision} for $I_\text{col}$.}
\end{figure}

We focus on the operator $I_\text{col} \left[ f,f \right]$ on the RHS of the Fokker-Planck equation \eqref{eq:Boltzmann}. In \ref{part:collision}, we assume a simplified model \eqref{eq:collision} for the sake of converging towards a closed set of equations, and also because there is no knowledge of what is the outcome of a collision between two jellyfish, not to mention the fact that the problem is reduced to a two-dimensional one on the context of this paper. The common observation in swarms is that collisions are not a factor explaining swarm formation, but rather a factor limiting aggregation, which is why we consider a model that has this property. However, for the sake of clarity and in order to build upon other active-matter models, we show in this appendix how Expression \eqref{eq:collision} can be seen as a mean-field approximation of a more general passive pairwise collision operator. Let us for instance consider
\begin{equation} \label{eq:Bertin_collision}
    \begin{split}
        I_\text{col}[f,f] &= -2 d_0 \mathcal{V} f(\theta) \int_{-\pi}^\pi \dd \theta_1 f(\theta_1) |\mathbf{e}_\theta - \mathbf{e}_{\theta_1}|\\[6pt]
        &+ 2 d_0 \mathcal{V} \int_{-\pi}^\pi \dd \theta_1 f(\theta_1) \int_{-\pi}^\pi \dd \theta_2 f(\theta_2) |\mathbf{e}_{\theta_1} - \mathbf{e}_{\theta_2}| \int_{-\infty}^\infty \dd \zeta p(\zeta) \sum_{m \in \mathbb{Z}} \delta (\theta_2 + \zeta - \theta + 2 m \pi) \ ,
    \end{split}
\end{equation}
where the dependency of $f$ and $\mathcal{V}$ to position $\mathbf{x}$ and time $t$ is implicit. The first term of Expression \eqref{eq:Bertin_collision} represents the amount of jellyfish with orientation $\theta$ prior to a collision event, changing orientation as a result of the collision occurring in the time interval $\dd t$, while the second term represents the amount of jellyfish whose orientation, $\theta_2$ prior to a collision event, becomes $\theta$ as result of a collision with an individual of orientation $\theta_1$ (Figure \ref{fig:process}c). As in \cite{bertin2006boltzmann}, only pairwise collisions are considered. Noting $d_0$ the size of jellyfish, and assuming that a collision occurs when two individuals become closer than $2 d_0$, then the number of jellyfish of orientation $\theta'$ colliding with a given one of orientation $\theta$ during the time interval $\dd t$ is given by $2 d_0 \mathcal{V} | \mathbf{e}_{\theta'} - \mathbf{e}_{\theta} | \dd t$ (Figure \ref{fig:collision}). Let us keep in mind that the model assumes that all jellyfish located at the same position are swimming at the same speed $\mathcal{V}$ (but possibly different orientations), which only depends on the local current speed (of course, jellyfish size $d_0$ is assumed negligible compared with the flow scale). Consistently, two jellyfish are swimming with speed $\mathcal{V}$ before the collision, and immediately readjust their speed to this value afterwards. Regarding reorientation after a collision, the model of \cite{gengel2024swarm} assumes soft repulsion interaction, however those cannot be straightforwardly translated into an angular collision operator, and besides there is no knowledge of whether jellyfish actually collide like soft spheres or not. In order to obtain a comprehensive framework without making strong assumptions about the nature of collisions themselves, we adopt the following model: after a collision, the two jellyfish exchange orientations, up to a random angle $\zeta$ distributed according to a Gaussian probability law of centre $0$ and standard deviation $\sigma$:
\begin{equation}
    p(\zeta) = \frac{1}{\sigma \sqrt{2 \pi}} \exp \left( -\frac{\zeta^2}{2 \sigma^2} \right) \ .
\end{equation}

Expression \eqref{eq:Bertin_collision} is thus similar to the operator used in \cite{bertin2006boltzmann} (Eq. (3)) for the Vicsek model, only here jellyfish are not trying to align with each other but rather nearly exchange orientations, like passive soft spheres colliding frontally. Like in \cite{bertin2006boltzmann}, it is insightful to project Expression \eqref{eq:Bertin_collision} in Fourier space, which is a necessary step to obtain hydrodynamic equations. Applying the Fourier transform to \eqref{eq:Bertin_collision} yields
\begin{equation} \label{eq:Icol_Fourier}
    \int_{-\pi}^\pi I_\text{col} \left[ f,f \right] \ee^{\ii n \theta} \, \dd \theta = -2 d_0 \mathcal{V} \left( 1 - \ee^{-n^2 \sigma^2/2} \right) \int_{-\pi}^\pi \dd \theta f(\theta) \ee^{\ii n \theta} \int_{-\pi}^\pi \dd \theta' f(\theta') |\mathbf{e}_\theta - \mathbf{e}_{\theta'}| \ .
\end{equation}

Using the relations
\begin{equation}
    f(\theta) = \frac{1}{2 \pi} \sum_{q \in \mathbb{Z}} f_q \ee^{-\ii q \theta} \quad \text{and} \quad |\mathbf{e}_\theta - \mathbf{e}_{\theta'}| = 2 \left| \sin\left( \frac{\theta - \theta'}{2} \right) \right| \ ,
\end{equation}
Expression \eqref{eq:Icol_Fourier} finally yields
\begin{equation} \label{eq:Icol_Fourier_final}
    \int_{-\pi}^\pi I_\text{col} \left[ f,f \right] \ee^{\ii n \theta} \, \dd \theta = \frac{8}{\pi} d_0 \mathcal{V} \left( 1 - \ee^{-n^2 \sigma^2/2} \right) \sum_{q \in \mathbb{Z}} \frac{f_q f_{n-q}}{4 q^2 - 1} \ .
\end{equation}

Expression \eqref{eq:Icol_Fourier_final} is still not convenient to use, therefore we simplify it further by assuming that $\sigma^2 / 2$ is small, and keeping only the term $q=0$ in the sum, which yields
\begin{equation} \label{eq:Icol_Fourier_simplified}
    \int_{-\pi}^\pi I_\text{col} \left[ f,f \right] \ee^{\ii n \theta} \, \dd \theta \equiv -\frac{4 \sigma^2}{\pi} d_0 \mathcal{V} \rho n^2 f_n \ .
\end{equation}

In this form, the collision operator is approximated by an additional angular diffusion term
\begin{equation}
    I_\text{col}[f,f] = \frac{4 \sigma^2}{\pi} d_0 \mathcal{V} \rho \partial_{\theta \theta} f \ ,
\end{equation}
which is the simplified model adopted in \ref{part:collision}.

\section*{Appendix B: Hydrodynamic closure for $\kappa \ll 1$}

In this appendix we derive the hydrodynamical equation in the limit of small $\kappa$, using the hypothesis that the Fourier modes $f_n$ of the distribution decrease with a power law, as in \cite{bertin2006boltzmann}. The purpose of this derivation is to demonstrate the consistency with Expressions \eqref{eq:continuity_unified} and \eqref{eq:complex_diffusion} derived in \ref{part:closure}, in the particular case of small $\kappa$ and $\epsilon$. Let us consider the Fokker-Planck equation \eqref{eq:Fokker-Planck} and project it in Fourier space by applying $\int_{-\pi}^\pi \dd \theta \ee^{\ii n \theta}$. The passive transport term in the LHS and diffusion-collision in the RHS straightforwardly yield
\begin{equation}
    \begin{split}
        \int_{-\pi}^\pi \dd \theta \, \ee^{\ii n \theta} \left( \partial_t f + \boldsymbol{\nabla} \cdot \left( f \mathbf{U} \right) \right) &= \partial_t f_n + \boldsymbol{\nabla} \cdot \left( f_n \mathbf{U} \right) \ ,\\[6pt]
        \text{and} \quad \int_{-\pi}^\pi \dd \theta \, \ee^{\ii n \theta} \mathcal{D}(\rho) \partial_{\theta \theta} f &= - n^2 \mathcal{D}(\rho) f_n \ .
    \end{split}
\end{equation}

The other terms can be treated using integration by parts and the definitions $\cos \theta = \left( \ee^{\ii \theta} + \ee^{-\ii \theta} \right)/2$ and $\sin \theta = \left( \ee^{\ii \theta} - \ee^{-\ii \theta} \right)/2\ii$:
\begin{equation}
    \begin{split}
        -\int_{-\pi}^\pi \dd \theta \, \ee^{\ii n \theta} \Psi f \sin \left( \theta - \theta_\Psi \right) &= \frac{n}{2} \underline{\Psi}^\star f_{n+1} - \frac{n}{2} \underline{\Psi} f_{n-1} \ ,\\[6pt]
        \text{and} \quad \int_{-\pi}^\pi \dd \theta \, \ee^{\ii n \theta} \boldsymbol{\nabla} \cdot \left( \mathcal{V} f \mathbf{e}_\theta \right) &= \frac{1}{2} \underline{\nabla}^\star \left( \mathcal{V} f_{n+1} \right) + \frac{1}{2} \underline{\nabla} \left( \mathcal{V} f_{n-1} \right) \ ,
    \end{split}
\end{equation}
where $\underline{\Psi} = \Psi_x + \ii \Psi_y = \Psi \ee^{\ii \theta_\Psi}$, $\underline{\nabla} = \partial_x + \ii \partial_y$, and $^\star$ is the complex conjugation. For all modes $n \in \mathbb{Z}$, we thus get an infinite set of coupled PDEs:
\begin{equation} \label{eq:Fourier_appendix}
    \partial_t f_n + \boldsymbol{\nabla} \cdot \left( f_n \mathbf{U} \right) + \frac{1}{2} \underline{\nabla} \left( f_{n-1} \mathcal{V} \right) + \frac{1}{2} \underline{\nabla}^\star \left( f_{n+1} \mathcal{V} \right) = -n^2 \mathcal{D}(\rho) f_n + \frac{n}{2} \underline{\Psi} f_{n-1} - \frac{n}{2} \underline{\Psi}^\star f_{n+1} \ .
\end{equation}

As introduced in \ref{part:polarisation}, Equation \eqref{eq:Fourier_appendix} with $n=0$ yields the continuity equation. In order to obtain an expression for the polarisation $\mathbf{p}$ (or, equivalently, $f_1$), we write \eqref{eq:Fourier_appendix} for $n=1$:
\begin{equation} \label{eq:Fourier_n=1}
    \partial_t f_1 + \boldsymbol{\nabla} \cdot \left( f_1 \mathbf{U} \right) + \frac{1}{2} \underline{\nabla} \left( \rho \mathcal{V} \right) + \frac{1}{2} \underline{\nabla}^\star \left( f_2 \mathcal{V} \right) = -\mathcal{D}(\rho) f_1 + \frac{1}{2} \underline{\Psi} \rho - \frac{1}{2} \underline{\Psi}^\star f_2 \ .
\end{equation}

In general, it is clear that this approach yields a recurrence relation for all modes $f_n$, however the Fourier expansion can be cut at $n=1$ if $\kappa$ and $\epsilon$ are both small (one can also expand to higher modes and notice the expected power law $f_n \sim \rho \kappa^{|n|}$). Indeed, in that case, \eqref{eq:Fourier_n=1} can be restricted to the dominating terms
\begin{equation}
    \frac{1}{2} \underline{\nabla} \left( \rho \mathcal{V} \right) \approx -\mathcal{D}(\rho) f_1 + \frac{1}{2} \underline{\Psi} \rho \ ,
\end{equation}
which yields the approximate expression
\begin{equation} \label{eq:p_limit_TT}
    \mathbf{p} \approx \frac{1}{2 \mathcal{D}(\rho)} \rho \boldsymbol{\Psi} - \frac{1}{2 \mathcal{D}(\rho)} \boldsymbol{\nabla} \left( \rho \mathcal{V} \right) \ .
\end{equation}

First of all, let us note that this expression is consistent with the polarisation of the von Mises distribution as written in Expression \eqref{eq:pvm_limit}. Indeed, the first term of Expression \eqref{eq:p_limit_TT}, which directly arises from steering, corresponds to $\mathbf{p}^\text{vM}$ in the limit $\kappa \ll 1$, while the second term, which is a diffusive transport, corresponds exactly to the fourth term of Expression \eqref{eq:complex_diffusion}, which is indeed the dominant one if $\kappa \ll 1$. In terms of angular distribution, cutting the Fourier sum at $n=\pm1$ means that the angular distribution is quasi-isotropic (since the steering rate is small compared to orientational diffusion), up to a small dipolar correction which arises mostly from steering:
\begin{equation}
    f(\theta) = \frac{\rho}{2 \pi} + \frac{1}{\pi} \mathbf{p} \cdot \mathbf{e}_\theta + \mathcal{O} (\rho \kappa^2) \ .
\end{equation}

\bibliography{bibliography}

\begin{thebibliography}{10}
\expandafter\ifx\csname url\endcsname\relax
  \def\url#1{\texttt{#1}}\fi
\expandafter\ifx\csname urlprefix\endcsname\relax\def\urlprefix{URL }\fi
\providecommand{\bibinfo}[2]{#2}
\providecommand{\eprint}[2][]{\url{#2}}

\bibitem{ouellette2022physics}
\bibinfo{author}{Ouellette, N.~T.}
\newblock \bibinfo{title}{A physics perspective on collective animal behavior}.
\newblock \emph{\bibinfo{journal}{Physical Biology}} \textbf{\bibinfo{volume}{19}}, \bibinfo{pages}{021004} (\bibinfo{year}{2022}).

\bibitem{douek2024long}
\bibinfo{author}{Douek, J.}, \bibinfo{author}{Giallongo, G.}, \bibinfo{author}{Harbuzov, Z.}, \bibinfo{author}{Galil, B.~S.} \& \bibinfo{author}{Rinkevich, B.}
\newblock \bibinfo{title}{Long-term population genetic features of the rhopilema nomadica jellyfish from the israeli mediterranean coasts}.
\newblock \emph{\bibinfo{journal}{Journal of Marine Science and Engineering}} \textbf{\bibinfo{volume}{12}}, \bibinfo{pages}{171} (\bibinfo{year}{2024}).

\bibitem{edelist2020phenological}
\bibinfo{author}{Edelist, D.} \emph{et~al.}
\newblock \bibinfo{title}{Phenological shift in swarming patterns of rhopilema nomadica in the eastern mediterranean sea}.
\newblock \emph{\bibinfo{journal}{Journal of Plankton Research}} \textbf{\bibinfo{volume}{42}}, \bibinfo{pages}{211--219} (\bibinfo{year}{2020}).

\bibitem{edelist2022tracking}
\bibinfo{author}{Edelist, D.} \emph{et~al.}
\newblock \bibinfo{title}{Tracking jellyfish swarm origins using a combined oceanographic-genetic-citizen science approach}.
\newblock \emph{\bibinfo{journal}{Frontiers in Marine Science}} \textbf{\bibinfo{volume}{9}}, \bibinfo{pages}{869619} (\bibinfo{year}{2022}).

\bibitem{johnson2001developing}
\bibinfo{author}{Johnson, D.~R.}, \bibinfo{author}{Perry, H.~M.} \& \bibinfo{author}{Burke, W.~D.}
\newblock \bibinfo{title}{Developing jellyfish strategy hypotheses using circulation models}.
\newblock \emph{\bibinfo{journal}{Hydrobiologia}} \textbf{\bibinfo{volume}{451}}, \bibinfo{pages}{213--221} (\bibinfo{year}{2001}).

\bibitem{vodopivec2025spatially}
\bibinfo{author}{Vodopivec, M.} \& \bibinfo{author}{Malej, A.}
\newblock \bibinfo{title}{Spatially explicit individual-based model reveals the mauve stinger jellyfish distribution in the mediterranean sea}.
\newblock \emph{\bibinfo{journal}{Ecological Modelling}} \textbf{\bibinfo{volume}{505}}, \bibinfo{pages}{111109} (\bibinfo{year}{2025}).

\bibitem{dehos2026short}
\bibinfo{author}{Dehos, S.} \emph{et~al.}
\newblock \bibinfo{title}{Short-range diversity and invasion dynamics of the freshwater jellyfish craspedacusta sowerbii in lake kinneret, israel, and its watershed: field sampling combined with hydrodynamical simulations}.
\newblock \emph{\bibinfo{journal}{NeoBiota}} \textbf{\bibinfo{volume}{107}}, \bibinfo{pages}{309--329} (\bibinfo{year}{2026}).

\bibitem{malul2019levantine}
\bibinfo{author}{Malul, D.}, \bibinfo{author}{Lotan, T.}, \bibinfo{author}{Makovsky, Y.}, \bibinfo{author}{Holzman, R.} \& \bibinfo{author}{Shavit, U.}
\newblock \bibinfo{title}{The levantine jellyfish rhopilema nomadica and rhizostoma pulmo swim faster against the flow than with the flow}.
\newblock \emph{\bibinfo{journal}{Scientific Reports}} \textbf{\bibinfo{volume}{9}}, \bibinfo{pages}{20337} (\bibinfo{year}{2019}).

\bibitem{albert2011s}
\bibinfo{author}{Albert, D.~J.}
\newblock \bibinfo{title}{What's on the mind of a jellyfish? a review of behavioural observations on aurelia sp. jellyfish}.
\newblock \emph{\bibinfo{journal}{Neuroscience \& Biobehavioral Reviews}} \textbf{\bibinfo{volume}{35}}, \bibinfo{pages}{474--482} (\bibinfo{year}{2011}).

\bibitem{fossette2015current}
\bibinfo{author}{Fossette, S.} \emph{et~al.}
\newblock \bibinfo{title}{Current-oriented swimming by jellyfish and its role in bloom maintenance}.
\newblock \emph{\bibinfo{journal}{Current Biology}} \textbf{\bibinfo{volume}{25}}, \bibinfo{pages}{342--347} (\bibinfo{year}{2015}).

\bibitem{malul2024directional}
\bibinfo{author}{Malul, D.} \emph{et~al.}
\newblock \bibinfo{title}{Directional swimming patterns in jellyfish aggregations}.
\newblock \emph{\bibinfo{journal}{Current Biology}} \textbf{\bibinfo{volume}{34}}, \bibinfo{pages}{4033--4038} (\bibinfo{year}{2024}).

\bibitem{marchetti2013hydrodynamics}
\bibinfo{author}{Marchetti, M.~C.} \emph{et~al.}
\newblock \bibinfo{title}{Hydrodynamics of soft active matter}.
\newblock \emph{\bibinfo{journal}{Reviews of Modern Physics}} \textbf{\bibinfo{volume}{85}}, \bibinfo{pages}{1143--1189} (\bibinfo{year}{2013}).

\bibitem{schnitzer1993theory}
\bibinfo{author}{Schnitzer, M.~J.}
\newblock \bibinfo{title}{Theory of continuum random walks and application to chemotaxis}.
\newblock \emph{\bibinfo{journal}{Physical Review E}} \textbf{\bibinfo{volume}{48}}, \bibinfo{pages}{2553} (\bibinfo{year}{1993}).

\bibitem{saragosti2012modeling}
\bibinfo{author}{Saragosti, J.}, \bibinfo{author}{Silberzan, P.} \& \bibinfo{author}{Buguin, A.}
\newblock \bibinfo{title}{Modeling e. coli tumbles by rotational diffusion. implications for chemotaxis}.
\newblock \emph{\bibinfo{journal}{PLOS ONE}} \textbf{\bibinfo{volume}{7}}, \bibinfo{pages}{e35412} (\bibinfo{year}{2012}).

\bibitem{liebchen2018synthetic}
\bibinfo{author}{Liebchen, B.} \& \bibinfo{author}{Lowen, H.}
\newblock \bibinfo{title}{Synthetic chemotaxis and collective behavior in active matter}.
\newblock \emph{\bibinfo{journal}{Accounts of Chemical Research}} \textbf{\bibinfo{volume}{51}}, \bibinfo{pages}{2982--2990} (\bibinfo{year}{2018}).

\bibitem{tjhung2018cluster}
\bibinfo{author}{Tjhung, E.}, \bibinfo{author}{Nardini, C.} \& \bibinfo{author}{Cates, M.~E.}
\newblock \bibinfo{title}{Cluster phases and bubbly phase separation in active fluids: reversal of the ostwald process}.
\newblock \emph{\bibinfo{journal}{Physical Review X}} \textbf{\bibinfo{volume}{8}}, \bibinfo{pages}{031080} (\bibinfo{year}{2018}).

\bibitem{toner1995long}
\bibinfo{author}{Toner, J.} \& \bibinfo{author}{Tu, Y.}
\newblock \bibinfo{title}{Long-range order in a two-dimensional dynamical xy model: how birds fly together}.
\newblock \emph{\bibinfo{journal}{Physical Review Letters}} \textbf{\bibinfo{volume}{75}}, \bibinfo{pages}{4326} (\bibinfo{year}{1995}).

\bibitem{bertin2006boltzmann}
\bibinfo{author}{Bertin, E.}, \bibinfo{author}{Droz, M.} \& \bibinfo{author}{Gr{\'e}goire, G.}
\newblock \bibinfo{title}{Boltzmann and hydrodynamic description for self-propelled particles}.
\newblock \emph{\bibinfo{journal}{Physical Review E}} \textbf{\bibinfo{volume}{74}}, \bibinfo{pages}{022101} (\bibinfo{year}{2006}).

\bibitem{cates2015motility}
\bibinfo{author}{Cates, M.~E.} \& \bibinfo{author}{Tailleur, J.}
\newblock \bibinfo{title}{Motility-induced phase separation}.
\newblock \emph{\bibinfo{journal}{Annual Review of Condensed Matter Physics}} \textbf{\bibinfo{volume}{6}}, \bibinfo{pages}{219--244} (\bibinfo{year}{2015}).

\bibitem{soto2024kinetic}
\bibinfo{author}{Soto, R.}, \bibinfo{author}{Pinto, M.} \& \bibinfo{author}{Brito, R.}
\newblock \bibinfo{title}{Kinetic theory of motility induced phase separation for active brownian particles}.
\newblock \emph{\bibinfo{journal}{Physical Review Letters}} \textbf{\bibinfo{volume}{132}}, \bibinfo{pages}{208301} (\bibinfo{year}{2024}).

\bibitem{gengel2023physics}
\bibinfo{author}{Gengel, E.}, \bibinfo{author}{Kuplik, Z.}, \bibinfo{author}{Angel, D.} \& \bibinfo{author}{Heifetz, E.}
\newblock \bibinfo{title}{A physics-based model of swarming jellyfish}.
\newblock \emph{\bibinfo{journal}{PLOS ONE}} \textbf{\bibinfo{volume}{18}}, \bibinfo{pages}{e0288378} (\bibinfo{year}{2023}).

\bibitem{gengel2024swarm}
\bibinfo{author}{Gengel, E.}, \bibinfo{author}{Kuplik, Z.}, \bibinfo{author}{Angel, D.} \& \bibinfo{author}{Heifetz, E.}
\newblock \bibinfo{title}{Swarm coherence mechanism for jellyfish}.
\newblock \emph{\bibinfo{journal}{Physical Review E}} \textbf{\bibinfo{volume}{110}}, \bibinfo{pages}{064406} (\bibinfo{year}{2024}).

\bibitem{gemmell2025movement}
\bibinfo{author}{Gemmell, B.~J.}, \bibinfo{author}{Colin, S.~P.} \& \bibinfo{author}{Costello, J.~H.}
\newblock \bibinfo{title}{Movement ecology of gelatinous zooplankton: approaches, challenges and future directions}.
\newblock \emph{\bibinfo{journal}{Journal of Experimental Biology}} \textbf{\bibinfo{volume}{228}}, \bibinfo{pages}{JEB247987} (\bibinfo{year}{2025}).

\bibitem{lushi2012collective}
\bibinfo{author}{Lushi, E.}, \bibinfo{author}{Goldstein, R.~E.} \& \bibinfo{author}{Shelley, M.~J.}
\newblock \bibinfo{title}{Collective chemotactic dynamics in the presence of self-generated fluid flows}.
\newblock \emph{\bibinfo{journal}{Physical Review E}} \textbf{\bibinfo{volume}{86}}, \bibinfo{pages}{040902} (\bibinfo{year}{2012}).

\bibitem{Kwok2009}
\bibinfo{author}{Kwok, R.}
\newblock \bibinfo{title}{Jellyfish help mix the world's oceans}.
\newblock \emph{\bibinfo{journal}{Nature}}  (\bibinfo{year}{2009}).

\bibitem{Katija2012}
\bibinfo{author}{Katija, K.}
\newblock \bibinfo{title}{Biogenic inputs to ocean mixing}.
\newblock \emph{\bibinfo{journal}{Journal of Experimental Biology}} \textbf{\bibinfo{volume}{215}}, \bibinfo{pages}{1040--1049} (\bibinfo{year}{2012}).

\bibitem{graham2001physical}
\bibinfo{author}{Graham, W.~M.}, \bibinfo{author}{Pag{\`e}s, F.} \& \bibinfo{author}{Hamner, W.~M.}
\newblock \bibinfo{title}{A physical context for gelatinous zooplankton aggregations: a review}.
\newblock \emph{\bibinfo{journal}{Hydrobiologia}} \textbf{\bibinfo{volume}{451}}, \bibinfo{pages}{199--212} (\bibinfo{year}{2001}).

\bibitem{rakow2006orientation}
\bibinfo{author}{Rakow, K.~C.} \& \bibinfo{author}{Graham, W.~M.}
\newblock \bibinfo{title}{Orientation and swimming mechanics by the scyphomedusa aurelia sp. in shear flow}.
\newblock \emph{\bibinfo{journal}{Limnology and Oceanography}} \textbf{\bibinfo{volume}{51}}, \bibinfo{pages}{1097--1106} (\bibinfo{year}{2006}).

\bibitem{alt1980biased}
\bibinfo{author}{Alt, W.}
\newblock \bibinfo{title}{Biased random walk models for chemotaxis and related diffusion approximations}.
\newblock \emph{\bibinfo{journal}{Journal of Mathematical Biology}} \textbf{\bibinfo{volume}{9}}, \bibinfo{pages}{147--177} (\bibinfo{year}{1980}).

\bibitem{manko2022oceanic}
\bibinfo{author}{Ma{\'n}ko, M.~K.}, \bibinfo{author}{Merchel, M.}, \bibinfo{author}{Kwasniewski, S.} \& \bibinfo{author}{Weydmann-Zwolicka, A.}
\newblock \bibinfo{title}{Oceanic fronts shape biodiversity of gelatinous zooplankton in the european arctic}.
\newblock \emph{\bibinfo{journal}{Frontiers in Marine Science}} \textbf{\bibinfo{volume}{9}}, \bibinfo{pages}{941025} (\bibinfo{year}{2022}).

\bibitem{kuramoto2003chemical}
\bibinfo{author}{Kuramoto, Y.}
\newblock \emph{\bibinfo{title}{Chemical oscillations, waves, and turbulence}} (\bibinfo{publisher}{Courier Corporation}, \bibinfo{year}{2003}).

\bibitem{vicsek1995novel}
\bibinfo{author}{Vicsek, T.}, \bibinfo{author}{Czir{\'o}k, A.}, \bibinfo{author}{Ben-Jacob, E.}, \bibinfo{author}{Cohen, I.} \& \bibinfo{author}{Shochet, O.}
\newblock \bibinfo{title}{Novel type of phase transition in a system of self-driven particles}.
\newblock \emph{\bibinfo{journal}{Physical Review Letters}} \textbf{\bibinfo{volume}{75}}, \bibinfo{pages}{1226} (\bibinfo{year}{1995}).

\bibitem{keller1970initiation}
\bibinfo{author}{Keller, E.~F.} \& \bibinfo{author}{Segel, L.~A.}
\newblock \bibinfo{title}{Initiation of slime mold aggregation viewed as an instability}.
\newblock \emph{\bibinfo{journal}{Journal of Theoretical Biology}} \textbf{\bibinfo{volume}{26}}, \bibinfo{pages}{399--415} (\bibinfo{year}{1970}).

\bibitem{keller1971model}
\bibinfo{author}{Keller, E.~F.} \& \bibinfo{author}{Segel, L.~A.}
\newblock \bibinfo{title}{Model for chemotaxis}.
\newblock \emph{\bibinfo{journal}{Journal of Theoretical Biology}} \textbf{\bibinfo{volume}{30}}, \bibinfo{pages}{225--234} (\bibinfo{year}{1971}).

\bibitem{oksendal2013stochastic}
\bibinfo{author}{Oksendal, B.}
\newblock \emph{\bibinfo{title}{Stochastic differential equations: an introduction with applications}} (\bibinfo{year}{2013}).

\bibitem{risken1989fokker}
\bibinfo{author}{Risken, H.}
\newblock \bibinfo{title}{Fokker-planck equation}.
\newblock In \emph{\bibinfo{booktitle}{The Fokker-Planck equation: methods of solution and applications}}, \bibinfo{pages}{63--95} (\bibinfo{year}{1989}).

\bibitem{gardiner2009stochastic}
\bibinfo{author}{Gardiner, C.}
\newblock \emph{\bibinfo{title}{Stochastic methods}}, vol.~\bibinfo{volume}{4} (\bibinfo{year}{2009}).

\bibitem{ernst2006boltzmann}
\bibinfo{author}{Ernst, M.}, \bibinfo{author}{Trizac, E.} \& \bibinfo{author}{Barrat, A.}
\newblock \bibinfo{title}{The boltzmann equation for driven systems of inelastic soft spheres}.
\newblock \emph{\bibinfo{journal}{Journal of statistical physics}} \textbf{\bibinfo{volume}{124}}, \bibinfo{pages}{549--586} (\bibinfo{year}{2006}).

\bibitem{gallagher2013newton}
\bibinfo{author}{Gallagher, I.}, \bibinfo{author}{Saint-Raymond, L.} \& \bibinfo{author}{Texier, B.}
\newblock \emph{\bibinfo{title}{From Newton to Boltzmann: hard spheres and short-range potentials}} (\bibinfo{publisher}{European Mathematical Society Z{\"u}rich}, \bibinfo{year}{2013}).

\bibitem{gimenez2025waterborne}
\bibinfo{author}{Gimenez, L.~H.}, \bibinfo{author}{Carroll, A.~R.} \& \bibinfo{author}{Pitt, K.~A.}
\newblock \bibinfo{title}{Waterborne conspecific cues enhance larval settlement in the invasive jellyfish aurelia coerulea}.
\newblock \emph{\bibinfo{journal}{Marine Biology}} \textbf{\bibinfo{volume}{172}}, \bibinfo{pages}{169} (\bibinfo{year}{2025}).

\bibitem{feliks2022intraseasonal}
\bibinfo{author}{Feliks, Y.}, \bibinfo{author}{Gildor, H.} \& \bibinfo{author}{Mantel, N.}
\newblock \bibinfo{title}{Intraseasonal oscillatory modes in the eastern mediterranean sea}.
\newblock \emph{\bibinfo{journal}{Journal of Physical Oceanography}} \textbf{\bibinfo{volume}{52}}, \bibinfo{pages}{1471--1482} (\bibinfo{year}{2022}).

\bibitem{mantel2024seasonal}
\bibinfo{author}{Mantel, N.} \emph{et~al.}
\newblock \bibinfo{title}{Seasonal and vertical tidal variability in the southeastern mediterranean sea}.
\newblock \emph{\bibinfo{journal}{Frontiers in Marine Science}} \textbf{\bibinfo{volume}{11}}, \bibinfo{pages}{1388137} (\bibinfo{year}{2024}).

\bibitem{von1918ganzzahligkeit}
\bibinfo{author}{von Mises, R.}
\newblock \bibinfo{title}{{\"U}ber die “ganzzahligkeit" der atomgewichte und verwandete fragen}.
\newblock \emph{\bibinfo{journal}{Physikalische Zeitschrift}} \textbf{\bibinfo{volume}{19}}, \bibinfo{pages}{490} (\bibinfo{year}{1918}).

\bibitem{silvano2026hydrodynamic}
\bibinfo{author}{Silvano, N.}, \bibinfo{author}{Hernandez-Garcia, E.} \& \bibinfo{author}{L{\'o}pez, C.}
\newblock \bibinfo{title}{Hydrodynamic description of proliferating active matter}.
\newblock \emph{\bibinfo{journal}{arXiv preprint arXiv:2608.03317}}  (\bibinfo{year}{2026}).

\bibitem{abramowitz1948handbook}
\bibinfo{author}{Abramowitz, M.} \& \bibinfo{author}{Stegun, I.~A.}
\newblock \emph{\bibinfo{title}{Handbook of mathematical functions with formulas, graphs, and mathematical tables}}, vol.~\bibinfo{volume}{55} (\bibinfo{publisher}{U.S. Government Printing Office}, \bibinfo{year}{1948}).

\bibitem{marshall1997finite}
\bibinfo{author}{Marshall, J.}, \bibinfo{author}{Adcroft, A.}, \bibinfo{author}{Hill, C.}, \bibinfo{author}{Perelman, L.} \& \bibinfo{author}{Heisey, C.}
\newblock \bibinfo{title}{A finite-volume, incompressible navier stokes model for studies of the ocean on parallel computers}.
\newblock \emph{\bibinfo{journal}{Journal of Geophysical Research: Oceans}} \textbf{\bibinfo{volume}{102}}, \bibinfo{pages}{5753--5766} (\bibinfo{year}{1997}).

\bibitem{kuplik2026rhopilema}
\bibinfo{author}{Kuplik, Z.} \emph{et~al.}
\newblock \bibinfo{title}{Rhopilema nomadica in the mediterranean: Molecular evidence for migration and insights into its proliferation}.
\newblock \emph{\bibinfo{journal}{Diversity}} \textbf{\bibinfo{volume}{18}}, \bibinfo{pages}{94} (\bibinfo{year}{2026}).

\bibitem{dror2024rising}
\bibinfo{author}{Dror, H.} \& \bibinfo{author}{Angel, D.}
\newblock \bibinfo{title}{Rising seawater temperatures affect the fitness of rhopilema nomadica polyps and podocysts and the expansion of this medusa into the western mediterranean}.
\newblock \emph{\bibinfo{journal}{Marine Ecology Progress Series}} \textbf{\bibinfo{volume}{728}}, \bibinfo{pages}{123--143} (\bibinfo{year}{2024}).

\bibitem{dror2025substrate}
\bibinfo{author}{Dror, H.} \& \bibinfo{author}{Angel, D.~L.}
\newblock \bibinfo{title}{Substrate preferences and the effect of temperature on planulae settlement of the scyphozoan rhopilema nomadica}.
\newblock \emph{\bibinfo{journal}{Hydrobiologia}} \bibinfo{pages}{1--14} (\bibinfo{year}{2025}).

\bibitem{shnapp2022myptv}
\bibinfo{author}{Shnapp, R.}
\newblock \bibinfo{title}{Myptv: a python package for 3d particle tracking}.
\newblock \emph{\bibinfo{journal}{Journal of Open Source Software}} \textbf{\bibinfo{volume}{7}}, \bibinfo{pages}{4398} (\bibinfo{year}{2022}).

\bibitem{diamant2023remote}
\bibinfo{author}{Diamant, R.}, \bibinfo{author}{Alexandri, T.}, \bibinfo{author}{Barak, N.} \& \bibinfo{author}{Lotan, T.}
\newblock \bibinfo{title}{A remote sensing approach for exploring the dynamics of jellyfish, relative to the water current}.
\newblock \emph{\bibinfo{journal}{Scientific Reports}} \textbf{\bibinfo{volume}{13}}, \bibinfo{pages}{14769} (\bibinfo{year}{2023}).

\bibitem{sauviat2025jellyfish}
\bibinfo{author}{Sauviat, A.}, \bibinfo{author}{Ponzo, Q.}, \bibinfo{author}{Bonnet, D.} \& \bibinfo{author}{Kerz{\'e}rho, V.}
\newblock \bibinfo{title}{Jellyfish journey live tracking using floating electronic tag}.
\newblock \emph{\bibinfo{journal}{Estuarine, Coastal and Shelf Science}} \textbf{\bibinfo{volume}{318}}, \bibinfo{pages}{109250} (\bibinfo{year}{2025}).

\end{thebibliography}

\end{document}